\documentclass[twocolumn]{aastex62}

\usepackage{natbib}
\usepackage{color}
\usepackage{verbatim}
\usepackage{rotating}
\usepackage{multirow}
\usepackage{url}
\usepackage{amsmath,amsfonts,amsthm,bm} 
\usepackage{hyperref}

\newcommand{\gan}{ExoGAN}

\graphicspath{{./}{figures/}}

\received{\today}
\revised{-}
\accepted{-}
\submitjournal{ApJ}

\shorttitle{\gan}
\shortauthors{Zingales \& Waldmann}

\begin{document}
	
\title{ExoGAN: Retrieving Exoplanetary Atmospheres \\ Using Deep Convolutional Generative Adversarial Networks}

\correspondingauthor{Tiziano Zingales}
\email{tiziano.zingales.15@ucl.ac.uk}
\author[0000-0001-6880-5356]{Tiziano Zingales}
\affiliation{University College London \\
	Gower St, Bloomsbury, London WC1E 6BT, United Kingdom}
\affiliation{INAF - Osservatorio Astronomico di Palermo\\
	Piazza del Parlamento 1, 90134 Palermo, Italy
}

\author[0000-0002-4205-5267]{Ingo P. Waldmann}
\affiliation{University College London \\
	Gower St, Bloomsbury, London WC1E 6BT, United Kingdom}





\begin{abstract}
	
	Atmospheric retrievals on exoplanets usually involve computationally intensive Bayesian sampling methods. Large parameter spaces and increasingly complex atmospheric models create a computational bottleneck forcing a trade-off between statistical sampling accuracy and model complexity. It is especially true for upcoming JWST and ARIEL observations.  
	We introduce \gan, the Exoplanet Generative Adversarial Network, a new deep learning algorithm able to recognise molecular features, atmospheric trace-gas abundances and planetary parameters using unsupervised learning. Once trained, \gan \ is widely applicable to a large number of instruments and planetary types. The \gan\ retrievals constitute a significant speed improvement over traditional retrievals and can be used either as a final atmospheric analysis or provide prior constraints to subsequent retrieval.

\end{abstract}

\section{Introduction}

The modelling of exoplanetary atmospheric spectroscopy through so-called atmospheric retrieval algorithms has become accepted standard in the interpretation of transmission and emission spectroscopic measurements \citep[e.g.][]{2018arXiv180500029K,2018AJ....155..156T,2018AJ....155...55B,2018arXiv180500424M,2018Natur.557...68S,2041-8205-850-2-L32,2017ApJ...834...50B,2016ApJ...833..120R}. These retrieval algorithms are designed to solving the often ill-posed inverse problem of determining atmospheric parameters (such as trace gas abundances for example) from the measured spectra and their corresponding measurement uncertainties \citep[e.g.][]{2008JQSRT.109.1136I,2009ApJ...707...24M,2013ApJ...775..137L,Benneke:2013vx,2017AJ....154...91L,doi:10.1093/mnras/stx2748,2016ascl.soft08004C}. The associated atmospheric forward model to be fitted varies in complexity from retrieval to retrieval but most times encompasses a high dimensional likelihood space to be sampled. In the era of JWST \citep{Gardner2006} and ARIEL \citep{2016SPIE.9904E..1XT} observations, said model complexity will have to increase significantly. To date, the most commonly adopted statistical sampling methods are Nested Sampling \citep{2004AIPC..735..395S, 2008MNRAS.384..449F, Feroz:2009jn} and Markov Chain Monte Carlo \citep[e.g.][]{gregory11}. These approaches typically require of the order of 10$^5$ - 10$^6$ forward model realisations until convergence. The traditional analysis method, which uses Bayesian statistics, creates a precarious bottleneck: to achieve convergence within reasonable time frames (hours to days), we require the atmospheric forward model to be fast and consequently overly simplistic. The inclusion of disequilibrium chemistry, self-consistent cloud models and the move from 1\,D to 2-3\,D radiative transfer, are largely precluded by this constraint. 
In this paper, we present the first deep learning architecture for exoplanetary atmospheric retrievals and discuss a path towards solving the computational bottleneck using atmospheric retrievals assisted by deep-learning. 

Artificial Intelligence has been used extensively to understand and describe complex structures and behaviour in a wide variety of dataset across a plethora of research fields. 

In recent years, the field of exoplanets has seen pioneering deep-learning papers on planet detection \citep{2018MNRAS.474..478P,2018AJ....155...94S}, exoplanet transit prediction \citep{2017MNRAS.465.3495K} and atmospheric spectral identification \citet{2016ApJ...820..107W}.
In \citet{2016ApJ...820..107W} we applied a deep-belief neural network (DBN) to recognise the atmospheric features of an exoplanetary emission spectrum. This approach provided a qualitative understanding of the atmospheric trace gases likely to be present in a planetary emission spectrum, to then be included in our atmospheric retrieval framework TauREx \citep{2015ApJ...802..107W,2015ApJ...813...13W}. In this paper, we introduce a generative adversarial network \citep[GAN,][]{2014arXiv1406.2661G} to predict the maximum likelihood (ML) of the full retrieval solution given the observed spectrum. As shown in the following sections, this can be used as a stand-alone solution to retrieval or used to constrain the prior parameter ranges for a more standard atmospheric retrieval later. 

We design our algorithm following four guiding principles: 

\begin{itemize}
	\item Once trained, the deep or machine-learning algorithm should apply to the widest possible range of planet types.
	\item Once trained, the algorithm should apply to a wide range of instruments.
	\item The algorithm should be robust in the presence of unknown `un-trained' features and be able to generalise to parameter regimes outside its formal training set.
	\item The design of the algorithm and data format should be modular and easily modifiable and expandable.
\end{itemize}

In the following sections, we present the Exoplanet Generative Adversarial Network (\gan) algorithm and demonstrate it on a variety of retrieval scenarios. We provide the \gan\ algorithm freely to the community (see end of paper).

\section{Method}

In the following sections, we will introduce GANs and deep convolutional generative adversarial networks (DCGANs), followed by a discussion how we adopt DCGANs for exoplanetary retrievals.

\subsection{Generative Adversarial Networks}

Generative Adversarial Networks first introduced by \citet{2014arXiv1406.2661G} belongs to the class of unsupervised deep generative neural networks \citep{Goodfellow-et-al-2016}. Deep generative models can learn the arbitrarily complex probability distribution of a data set, $p_{data}$, and can generate new data sets drawn from $p_{data}$. Similarly, they can also be used to fill in missing information in an incomplete data set, so-called inpainting. In this work, we use the data inpainting properties of the GAN to perform retrievals of the atmospheric forward model parameters. 

The most common analogy for a GAN architecture is that of a counterfeit operation. The neural network is given a training data set, $\bf{x}$, in our case combinations of atmospheric spectra with their associated forward model parameters. We refer to the training set as the `real' data with the probability distribution $p_{data}$. Now two deep neural networks are pitted against each other in a  \texttt{minmax} game. One network, the generator network ($G$), will try to create a `fake' dataset ($p_{g}$), indistinguishable from the `real' data. In a second step, a second neural network, the discriminator ($D$), tries to classify `fake' from `real' data correctly. The training phase of the GAN is completed when a Nash equilibrium is reached, and the discriminator cannot identify real from fake any longer. At this stage the generator network will have learned a good representation of the data probability distribution and $p_{g} \simeq p_{data}$. Figure~\ref{fig:gan_schema} shows a schematic of our GAN implementation. Unlike for variational inference methods, such as variational autoencoders \citep[VAE;][]{2013arXiv1312.6114K,2014arXiv1401.4082J}, the functional form of the data likelihood does not need to be specified but is learned by the Generator. Such implicit latent variable models or likelihood-free networks allow the learning of arbitrarily complex probability distributions in an unsupervised manner while assuming minimal prior assumptions on the data distribution. 

GANs have been applied to multiple problems, such as semi-supervised learning, stabilizing sequence learning methods for speech and language, and 3D modelling \citep{2015arXiv150605751D, 2015arXiv151106434R, 2016arXiv160603498S, 2016arXiv160203220L, 2016arXiv160908144W}.
Notable examples of GANs applied in an astrophysical context are given by \citet{2018arXiv180109070R,2018MNRAS.477.2513S,doi:10.1093/mnrasl/slx008}, who used GANs trained on existing N-body simulations to efficiently generate new, physically realistic realisations of the cosmic web, learn Point Spread Function from data or de-noise ground-based observations of galaxies. 

In the field of exoplanets, the use of GANs or similar deep architectures has not yet been explored. In this work, we base \gan\ on a Deep Convolutional Generative Adversarial Network \citep[DCGAN,][]{2015arXiv151106434R}. 

DCGANs are an evolution from the classical GAN by replacing the multilayer perceptrons \citep[MLPs;][]{Rumelhart:1986:LIR:104279.104293,Bengio:2009kb} in the Generator and Discriminator networks with all convolutional layers. Their characteristics makes DCGAN significantly more robust to discrete-mode and manifold model collapse  \citep{2016arXiv161102163M,2017arXiv170104862A} and are found to be stable in most training scenarios \citep{2015arXiv151106434R}. The use of batch normalisation (appendix~\ref{app:bn} further increases training speed and robustness. Besides, we note that convolutional networks are ideally suited to capturing the highly correlated signals of broad, roto-vibrational spectral bands in NIR and IR wavelengths. 

\subsection{Adversarial Training}

\begin{figure*}
	\centering
	\includegraphics[width=\textwidth]{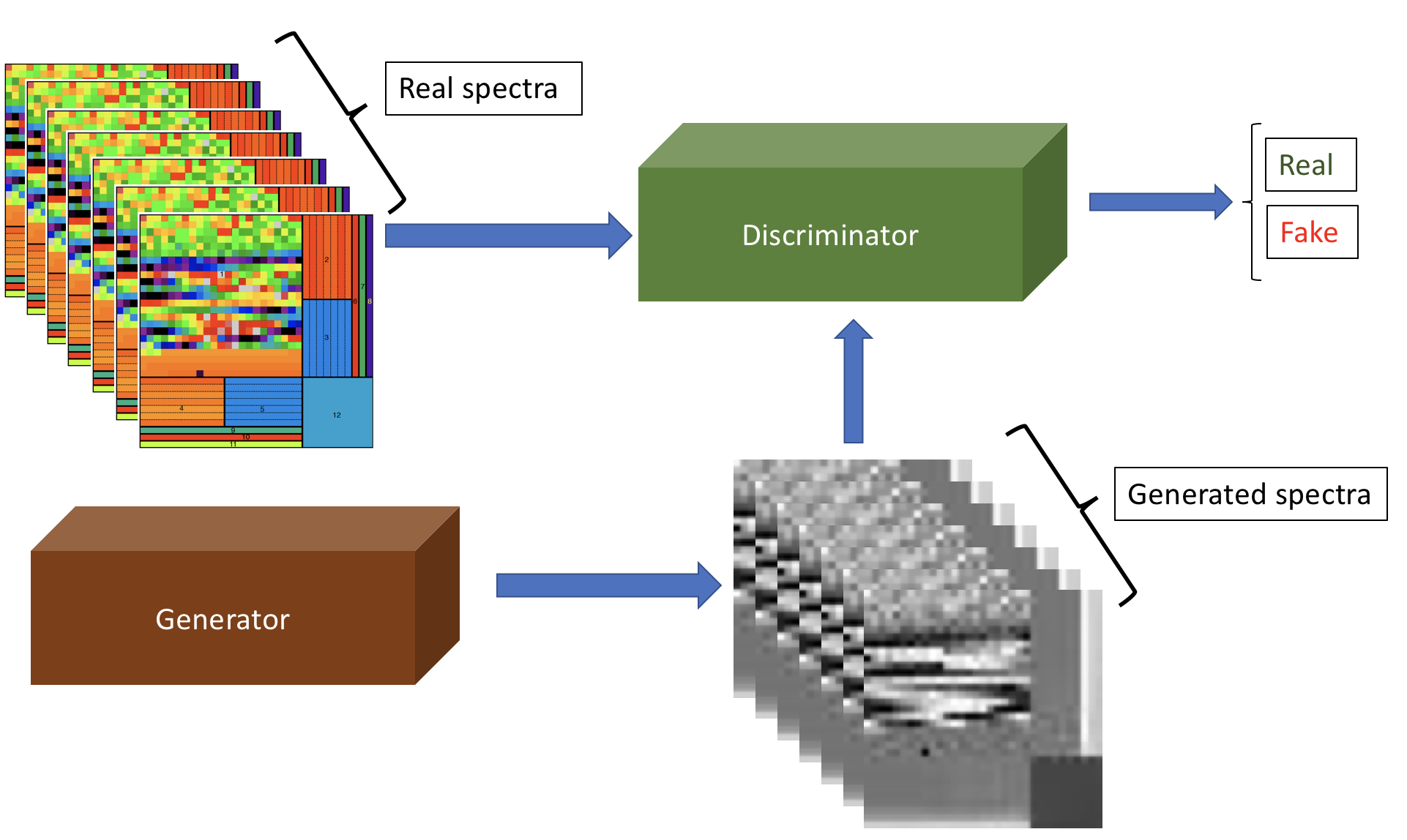}
	\caption{The \gan\ scheme. The Generator produces datasets sampling from a latent variable space ${\bf z}$. The Discriminator compares the generated dataset with data drawn from the training set (top left). The network has converged when the Discriminator cannot differentiate Real spectra from Generated Spectra any longer. \label{fig:gan_schema}}
\end{figure*}

As described in the previous section, both Generator and Discriminator networks are pitted against one another during training. The goal of the training phase is to reach a Nash Equilibrium, i.e. when neither player can improve by unilaterally changing one's strategy.  Figure~\ref{fig:gan_schema} shows a schematic of the \gan~setup.

In order to return the generator distribution $p_g$ over the data $\textbf{x}$ we start from a prior distribution of Gaussian distributed latent variables $p(\textbf{z})$ and define $G(\textbf{z};{\bm \theta}_G)$ as the mapping from latent variable space to generated data. Here ${\bm \theta}_G$ are the hyperparameters of the Generator network (see table \ref{tab:gan_architecture}). 

\noindent Let $D(\textbf{x})$ be the probability that $\textbf{x}$ came from the data rather than $p_g$. Hence, in the state of convergence, we have $p_{g} = p_{data}$ and $D(\bf{x}) = \frac{1}{2}$. In the training phase we need $D$ to maximise the probability of assigning the correct label to both training examples and samples from $G$. At the same time we want $G$ to minimize the probability $\log{\left(1 - D(G(\textbf{z}))\right)}$. We can now define the cross-entropy cost-function of the Discriminator as:

\begin{equation}
J^{(D)}   = - \left[ \log D({\bf x}) + \log\left(1 - D\left(G({\bf z})\right) \right) \right]
\label{eq:binary_cross_entropy3}
\end{equation}

\noindent During training, we employ batch training, with the cost function of a batch of $n$ data samples being 
\begin{equation}
\begin{aligned}
J^{(D)}  = & - \left\lbrace \sum_{i=1}^n \log D(x_i) + \right.\\
& + \left. \sum_{i=1}^n \log \left( 1 - D(G(z_i)) \right)\right\rbrace
\label{eq:binary_batches}
\end{aligned}
\end{equation}

\noindent which can be written as the expectation values over the data and generated samples:

\begin{align}
\label{eq:J_function}
J^{(D)} & = - \lbrace \mathbb{E}_{\textbf{x}\sim p_{data}}[\log D({\bf x})] \\\nonumber
& + \mathbb{E}_{\textbf{z}\sim p_{z}}[\log \left( 1 - D(G({\bf z}))\right)]\rbrace.
\end{align}

\noindent Since the discriminator wants to minimize the cost function and the generator wants to maximise it, we can summarise the training as a zero-sum game where the cost function for the generator is given by: $J^{(G)} = - J^{(D)}$. Hence, to capture the entire game, we only need to specify the loss-function of the Discriminator since it encompasses both ${\bm \theta}^{(D)}$ and  ${\bm \theta}^{(G)}$ hyperparameters. We then optimise the   value function $V  ({\bm \theta}^{(D)}, {\bm \theta}^{(G)}) = - J^{(D)} ({\bm \theta}^{(D)}, {\bm \theta}^{(G)})$, 

\begin{align}
\label{eq:min_max}
\min_G \max_D V(D, G) & = \mathbb{E}_{\textbf{x}\sim p_{data}}[\log D({\bf x})] \\\nonumber
& + \mathbb{E}_{\textbf{z}\sim p_{z}}[\log \left( 1 - D(G({\bf z}))\right)].
\end{align}

\noindent As stated earlier, equation~\ref{eq:min_max} constitutes a $\mathtt{minmax}$ game since it involves minimising over $G$ in an outer loop and maximising over $D$ in an inner loop.

\begin{figure*}
	\centering
	\includegraphics[width=\textwidth]{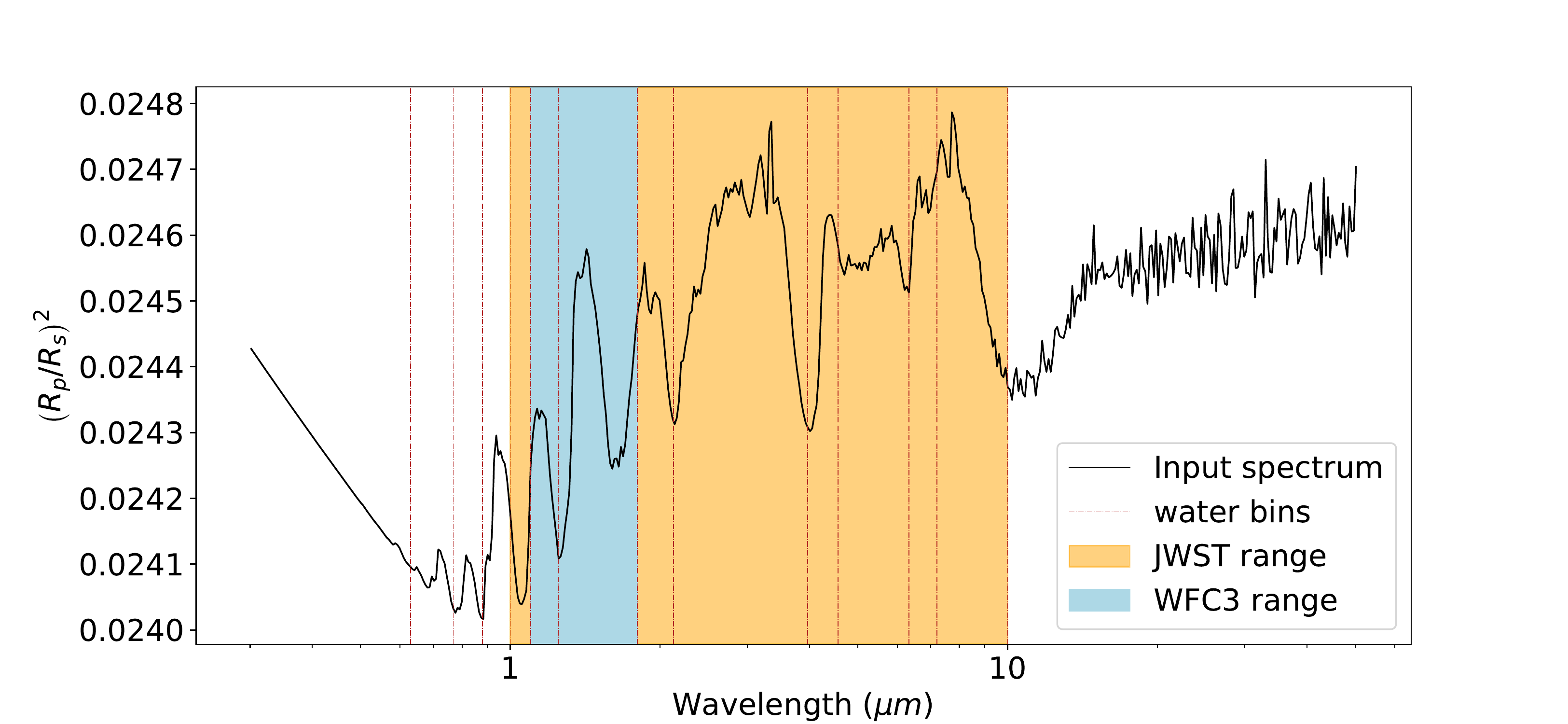}
	\caption{Spectral binning used in this work. The black line is a simulated spectrum of the hot-Jupiter HD\,189733b. The red vertical lines represent the bin edges of prominent water bands. The blue and orange areas are the Hubble/WFC3 and JWST band-passes considered in this paper, respectively .\label{fig:binned_spectrum}}
\end{figure*}

\subsection{Application to exoplanet spectra}

Here we explain the data format of the input and training data. In figure~\ref{fig:binned_spectrum} we show an example a transmission spectrum of a cloud-free hot-Jupiter with water as the only trace-gas at $3\cdot 10^{-4}$ volume mixing ratio at a constant resolution of $\frac{\Delta \lambda}{\lambda} = 100$. We train \gan~on a wavelength range of $0.3\mu$m $- 50 \mu$m. For this paper, we restrict our sampling resolution to be R = 100 for every spectrum. This choice, however, does not preclude training with higher resolution data in the future. 

\subsubsection{Normalisation}

For the neural network to learn efficiently, we must normalise the data to lie between zero and unity. We have experimented with various normalisation schemes. The most obvious scheme is a `global' normalisation, where we normalise the full training set by its global maximum and minimum values. This approach proved problematic as spectral signatures for planets with low trace-gas abundances, and small atmospheric scale heights, would be too weak/flat to be recognisable by the neural network for reasonable training times. We have therefore opted to normalise each training spectrum to amplify the spectral features. Assuming that the most common broadband absorber is water in an exoplanetary atmosphere, we divide the spectral range along its major water bands in the IR, see dashed red lines in Fig \ref{fig:binned_spectrum}. Note this does not mean that water-free atmospheres cannot be detected. Additionally, we divide the spectrum by the pass-bands of the JWST/NIRISS, NIRCam and MIRI instruments \citep{doi:10.1080/00107514.2018.1467648} and the Hubble/WFC3 instrument passband. In total, we have 14 spectral bands. We now normalise each spectral band between 0 and 1 and record the minimum and maximum normalisation factors for each. This normalisation scheme ensures a maximum amplification of the spectral features while retaining reversibility. 

\subsubsection{The Atmospheric Spectrum and Parameters Array (ASPA)}

\begin{figure}
	\centering
	\includegraphics[width=0.4\textwidth]{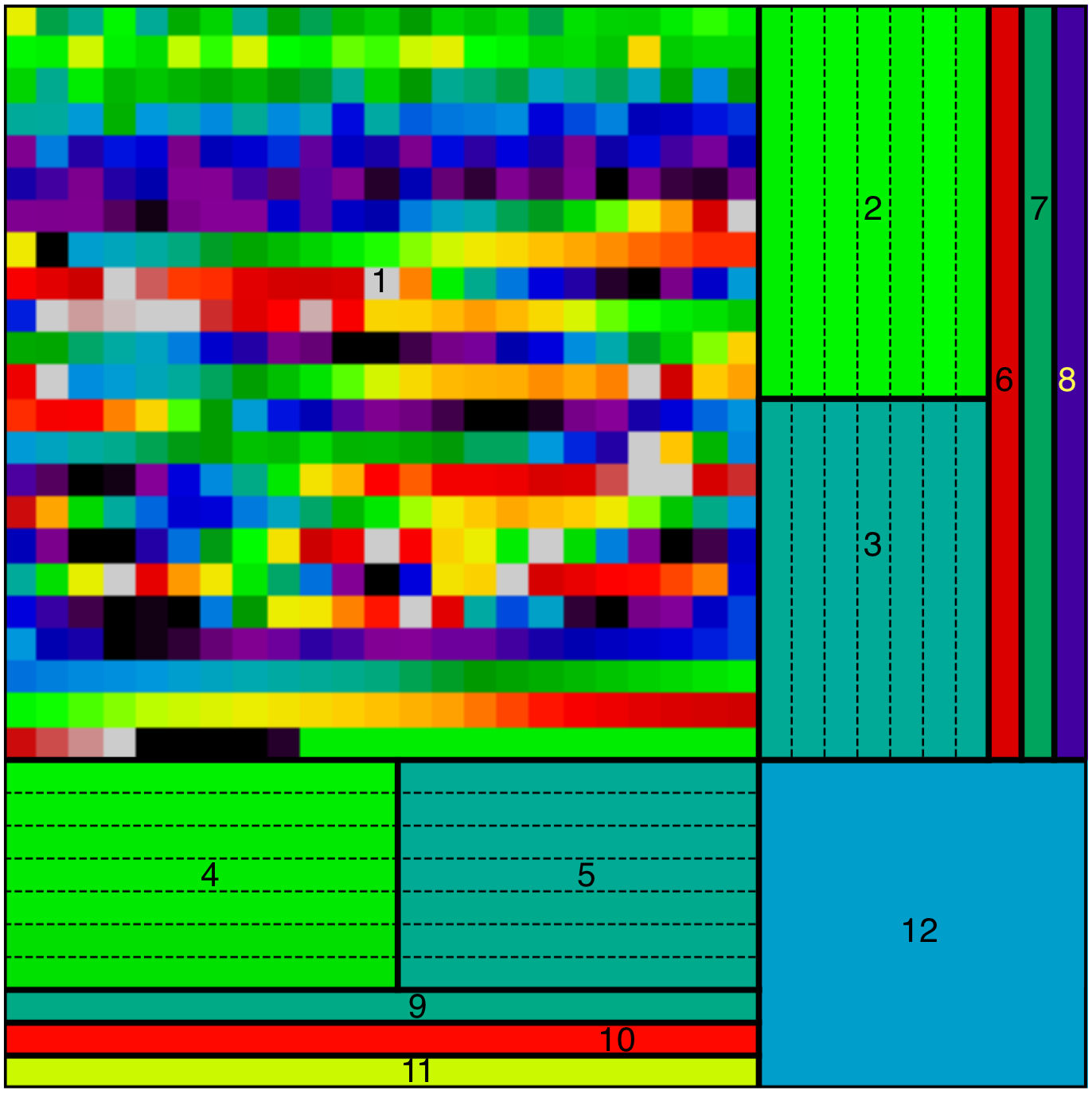}
	\caption{The Atmospheric Spectra and Parameters Array (ASPA). Each area is dedicated to a particular atmospheric characteristic: Area \textbf{1} is the spectrum between $1 \mu$m and $50 \mu$m at resolution 100 normalised between 0 and 1 in each spectral bin. Areas \textbf{2} to \textbf{5} give information about the normalisation factors used in the different section of the spectrum, clear and dark area give, respectively, information about the maximum values and the minimum values. In areas \textbf{6} to \textbf{8} we encode the atmospheric trace-gas volume mixing ratios of CO$_2$, CO and CH$_4$ respectively. Areas \textbf{9} to \textbf{11} are, respectively $M_p$, $R_p$ and $T_p$. Area \textbf{12} gives information on the H$_2$O trace-gas volume mixing ratio.\label{fig:training_norm}}
\end{figure}

To store all aspects of an atmospheric transmission spectrum, we define the Atmospheric Spectrum and Parameters Array (ASPA). It is a 2D array encoding the 1D normalised spectral bands, each band's minimum and maximum normalisation factors and the associated forward model parameter values. We parametrise each training spectrum with seven forward model parameters, ${\bm \phi}$, namely: H$_2$O, CO$_2$, CH$_4$ and CO volume mixing ratios, the mass of the planet $M_p$, the radius $R_p$ and its isothermal temperature $T_p$ at the terminator.
Figure~\ref{fig:training_norm} shows a false-colour ASPA. For this paper, the ASPA is a 33$\times$33 pixel array, with the main part (section 1) encoding the spectral information. Sections 2 - 5 encode the normalisation factors and 6 - 12 the atmospheric parameters. By design, the planet's water abundance takes a significantly large range area of the ASPA, reflecting the relative importance of water in forming the spectral continuum. The ASPA format is adaptable to other configurations in the future.

\subsection{The training}
To train \gan~on a wide range of possible exoplanetary atmospheres, we generated a very comprehensive training set of atmospheric forward models using the TauREx retrieval code \citep{2015ApJ...813...13W, 2015ApJ...802..107W}. We sampled each of the seven previously mentioned forward model parameters (H$_2$O, CO$_2$, CH$_4$ and CO abundances, the mass of the planet $M_p$, the radius $R_p$ and the temperature $T_p$) 10 times within the parameter ranges denoted in table~\ref{tab:par_bounds}. This configuration yields 10$^7$ forward models, which are split into 90\% training set and 10\% test set. The test set is used to validate the accuracy of the network on previously unseen data. As discussed later on, we find this training set to be overcomplete and only require a smaller subset of the full training set for convergence. 

During the training, we perform two training iterations of the discriminator to every training step of the generator. We used an NVIDIA TESLA V100 GPU with minibatch sizes of 64 training ASPAs. We required $\sim 9$ hours per epoch on the V100 GPU and comparatively about three days on 20 CPU cores in parallel. The convergences of the loss functions during the training phase are shown in figure \ref{fig:training_losses}.
The full model setup can be found in the appendix (table \ref{tab:gan_params}). We tested three different sizes of our latent variable space ${\bf z}$, with  
$z_{dim}$ = 50, 100 and 200. We found  $z_{dim} = 50$, to yield significantly noisier reconstructions at the end of one epoch of training, whereas no discernible differences between $z_{dim} = 100$ and $z_{dim} = 200$ could be observed. We hence settled on $z_{dim} = 100$. We have adopted a training minibatch size of 64 ASPAs and found no significant effect of larger training batch sizes on network convergence.

During minibatch training, the algorithm is presented with a sub-set of the full training data (in this case 64 ASPAs) rather than the full training set (or batch). This eases memory requirements of large training set, in particular for memory limited devices such as GPUs. By only considering a sub-set of training data at a time, a gradient descent optimiser, such as ADAM, is still able to perform well, despite the increase in variance on the gradient estimated. In order to avoid biased estimations and convergence to local minima, minibatches must be selected randomly from the training set at each iteration.

\begin{table}
	\centering
	\begin{tabular}{ |p{2cm}||p{2cm}|p{2cm}|  }
		\hline
		\multicolumn{3}{|c|}{Training set parameters} \\
		\hline
		Variable & lower bound & upper bound \\
		\hline
		H$_2$O    & $10^{-8}$    & $10^{-1}$    \\
		CO$_2$    & $10^{-8}$    & $10^{-1}$    \\
		CO    & $10^{-8}$    & $10^{-1}$    \\
		CH$_4$    & $10^{-8}$    & $10^{-1}$    \\
		$M_p$    & $0.8$ $\text{M}_J$    & $2.0$ $\text{M}_J$    \\
		$R_p$    & $0.8$ $\text{R}_J$    & $1.5$ $\text{R}_J$    \\
		$T_p$    & $1000$ K& $2000$ K    \\
		\hline
	\end{tabular}
	\caption{Parameters boundary condition used to generate the training set. Each parameter has been divided into 10 parts and used to model $10^7$ different spectra.\label{tab:par_bounds}}
\end{table}

\begin{figure}
	\centering
	\includegraphics[width=0.5\textwidth]{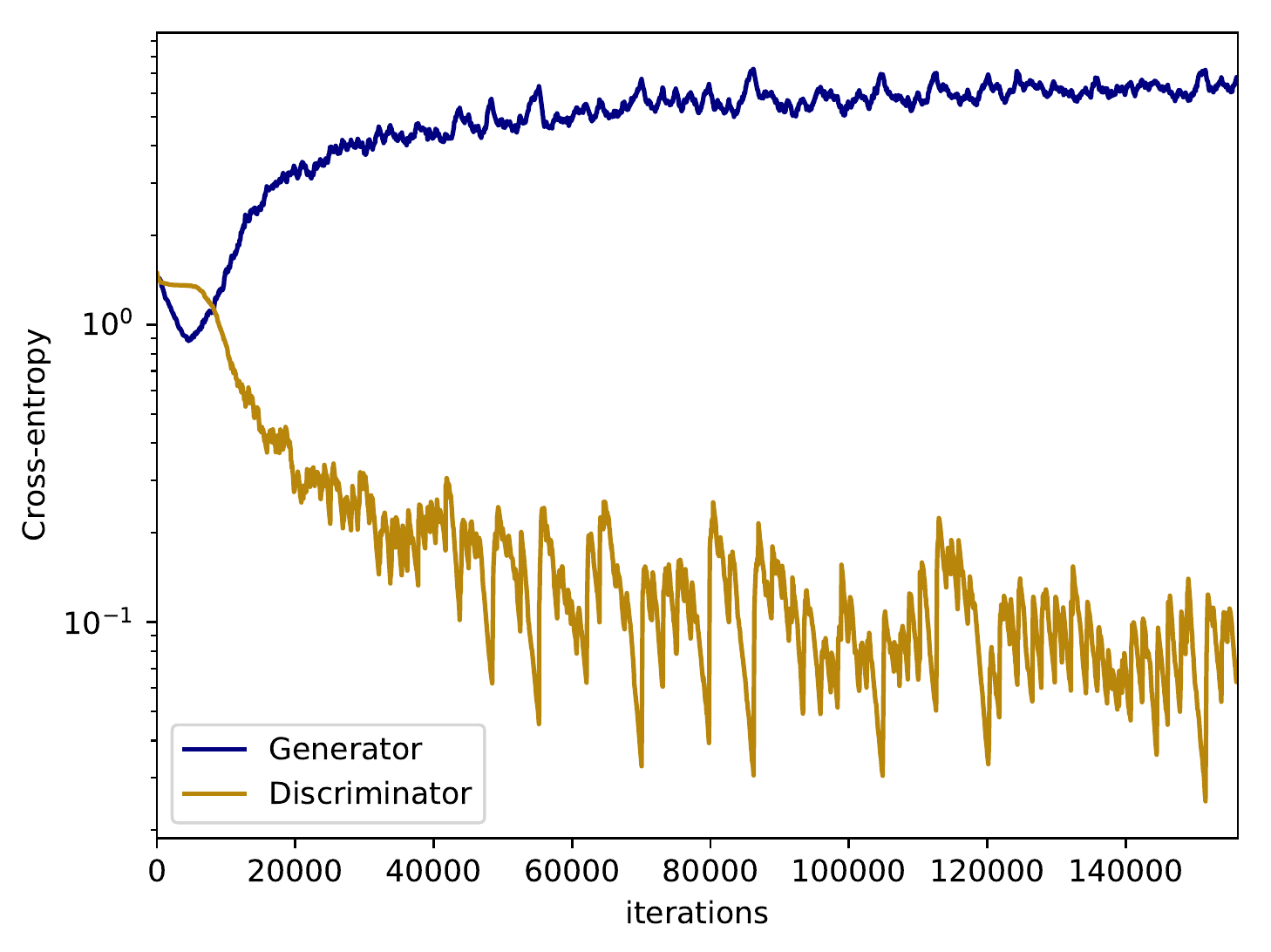}
	\caption{Discriminator \textbf{(golden)} and Generator \textbf{(blue)} cross-entropies as function of the iteration steps. \label{fig:training_losses}}
\end{figure}

\subsection{Data reconstruction}

Once we have trained \gan, we can now define our `retrieval' model. As alluded to above, we use the inpainting properties of a GAN to complete the missing data, in this case, the forward model parameters, in our ASPA. In other words, we convert our observed spectrum into the ASPA format and keep unknown values (parameters and missing wavelength ranges) masked. Given the information available, the \gan~will then attempt to fill in the missing information to complete the full ASPA. Here we follow the semantic inpainting algorithm by \citet{2016arXiv160707539Y}. 

We can define our reconstructed data, $ {\bf x}_{recon}$, from the incomplete observed data, ${\bf y}$, using 

\begin{equation}
{\bf x}_{recon} = M \odot {\bf y} + (1 - M) \odot G(\hat{\bf z})
\label{eq:im_reconstructed}
\end{equation}

\noindent where $M$ is a binary mask set to zero for missing values in ${\bf y}$, i.e. forward model parameter values and, possibly, missing wavelength ranges. Here, $\odot$ constitutes the Hadamard product and $G(\hat{\bf z})$ is the GAN generated data. We note that after the \gan~has been trained, ${\bf z}$ represents an encoding manifold of $p_{data}$ and we denote the closest match of $(M \odot G({\bf z}))$ to $(M \odot {\bf y})$ with $\hat{\bf z}$, where $\hat{\bf z} \subseteq {\bf z}$. 
The aim is now to obtain $\hat{\bf z}$ that accurately completes ${\bf x}_{recon}$.

Let us define the following optimisation.

\begin{equation}
\hat{\bf z} = \text{arg} \min_z \mathcal{L}({\bf z}).
\label{eq:z_hat}
\end{equation}

\noindent where $\mathcal{L}$ is a loss function of ${\bf z}$ that finds its minimum when $\hat{\bf z}$ is reached. Following  \citet{2016arXiv160707539Y}, we define the loss function to be comprised of two parts, contextual loss and perceptual loss,

\begin{equation}
\mathcal{L} = \mathcal{L}_{cont}({\bf z}) + \lambda \mathcal{L}_{perc}({\bf z}).
\label{eq:complete_loss}
\end{equation}

The contextual loss, $\mathcal{L}_{cont}({\bf z})$ is the difference between the observed data and the generated data. Here we follow the definition by \citet{amos2016image}:

\begin{equation}
\mathcal{L}_{cont}({\bf z}) = \parallel M \odot G({\bf z}) - M \odot y \parallel_1 .
\label{eq:contextual_loss}
\end{equation}

Empirically, \citet{2016arXiv160707539Y} find the $l_1$ norm to yield slightly better results, though the $l_2$ norm can equally be used. 
Whereas the conceptual loss compares the generated data with the observed data directly, the perceptual loss, $\mathcal{L}_{perc}({\bf z})$, uses the discriminator network to verify the validity of the generated data given the training set.

\begin{equation}
\mathcal{L}_{perc}({\bf z}) = \log \left( 1 - D(G({\bf z}))\right)
\label{eq:perceptual_loss}
\end{equation}

To solve equation~\ref{eq:z_hat} we use the ADAM optimiser \citep{2014arXiv1412.6980K} with a learning rate of 0.1. For a deeper discussion about the ADAM optimiser, see Appendix \ref{app:adam}.

We investigated the ratio of perceptual loss (Eq \ref{eq:perceptual_loss}) to contextual loss (Eq \ref{eq:contextual_loss}) and found $\lambda = 0.1$ to be optimal but note that $\lambda > 0.1$ gives too much emphasis to the perceptual loss term and yielded less reliable results.  

\begin{figure*}
	\centering
	\includegraphics[width=0.3\textwidth]{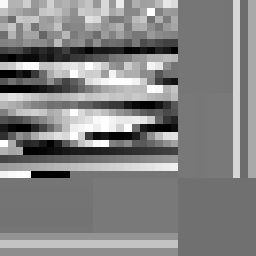}
	\includegraphics[width=0.3\textwidth]{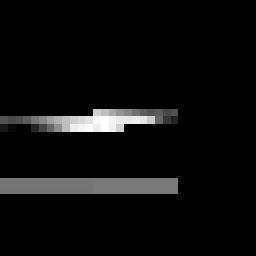}
	\includegraphics[width=0.3\textwidth]{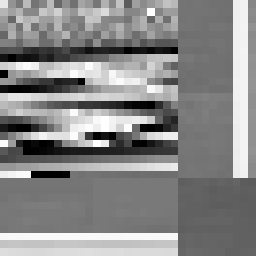}
	\caption{\textbf{Left}: input spectrum together with the parameters pixels. \textbf{Centre}: masked ASPA leaving Hubble/WFC3 wavelengths only. \textbf{Right}: \gan\ completed ASPA given the middle ASPA. \label{fig:predict_phase_wfc3}}
\end{figure*}

\begin{figure*}
	\centering
	\includegraphics[width=0.3\textwidth]{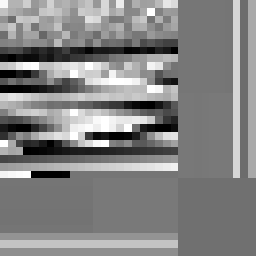}
	\includegraphics[width=0.3\textwidth]{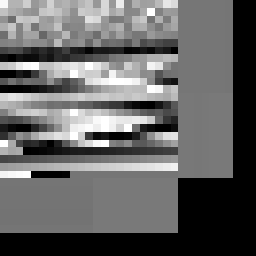}
	\includegraphics[width=0.3\textwidth]{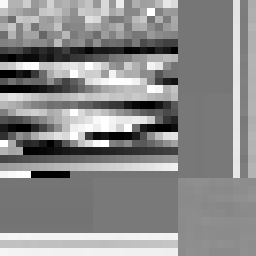}
	\caption{Same as figure~\ref{fig:predict_phase_wfc3} but only masking the atmospheric forward model parameters. \label{fig:predict_phase_par}}
\end{figure*}

\begin{figure*}
	\centering
	\includegraphics[width=0.7\textwidth]{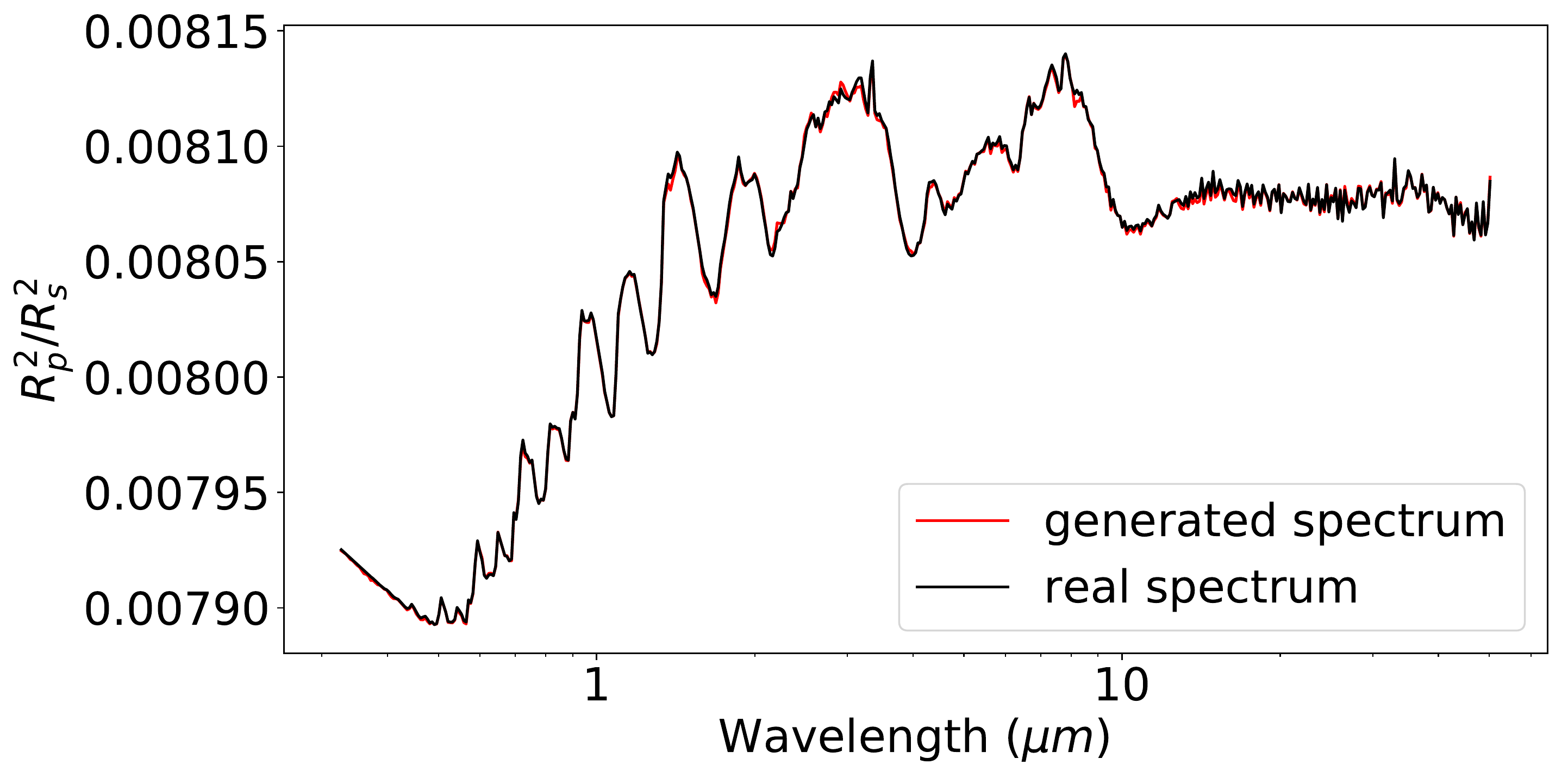}
	\caption{Spectral reconstruction of \gan of a water dominated Hubble/WFC3 spectrum. Black: the ground-truth spectrum; Red: the \gan\ reconstructed spectrum across all wavelengths giving as input only the Hubble/WFC3 band-pass. \label{fig:reconstructed_spectrum}}
\end{figure*}

In figures \ref{fig:predict_phase_wfc3}~\&~\ref{fig:predict_phase_par} we show the three phases associated to a prediction: Left, the ground truth; Middle: the masked spectrum/parameters; Right: the reconstructed ASPA. 
Figure \ref{fig:reconstructed_spectrum} shows 
a water-dominated atmosphere of a test-set hot-Jupiter (black) and the \gan\ reconstructed spectrum based on the Hubble/WFC3 bandpass only (red). We find a very good agreement between reconstructed and ground-truth spectra.

\section{Atmospheric parameter retrieval}

To retrieve the atmospheric forward model parameters, we assume the observational uncertainties on the spectrum to be Gaussian distributed. We then generate 1000 noisy instances of the observed spectrum, $x_i(\lambda)$, by sampling from a normal distribution with a mean of $x(\lambda)$ and standard deviation $\sigma_\lambda$. From these noisy spectrum instances, we generate 1000 corresponding ASPAs with missing information (may they be parameters, spectral ranges or both) masked. We now let \gan\ predict and inpaint these ASPAs. Finally, we collect all parameter predictions and calculate the mean and standard deviation of the resulting distribution. Hence, the resulting distributions are not posterior distributions derived from a Nested or MCMC sampling atmospheric retrieval, but are conceptually more similar to running a retrieval based on optimal-estimation multiple times and collecting the distribution of results.

\section{Accuracy tests}

We defined the accuracy of the retrieved parameter, $A$, as the function of the ground-truth parameter value, $\phi$, the retrieved  value, $\phi_{recon}$, and its corresponding error $\sigma_{\phi}$, 

\begin{equation}
A(\phi,\sigma_\phi)= \frac{1}{N} \sum_i^N \frac{({\bm \phi}_{i,recon}  - {\bm \phi_{i}})^2}{{\bm \phi_{i}}^2 + {\sigma}_{\phi_{i}}^2}
\label{eq:accuracy}
\end{equation}

\noindent where $N$ is the number of reconstructed ASPA instances. 

\begin{figure*}
	\centering
	\includegraphics[width=\textwidth]{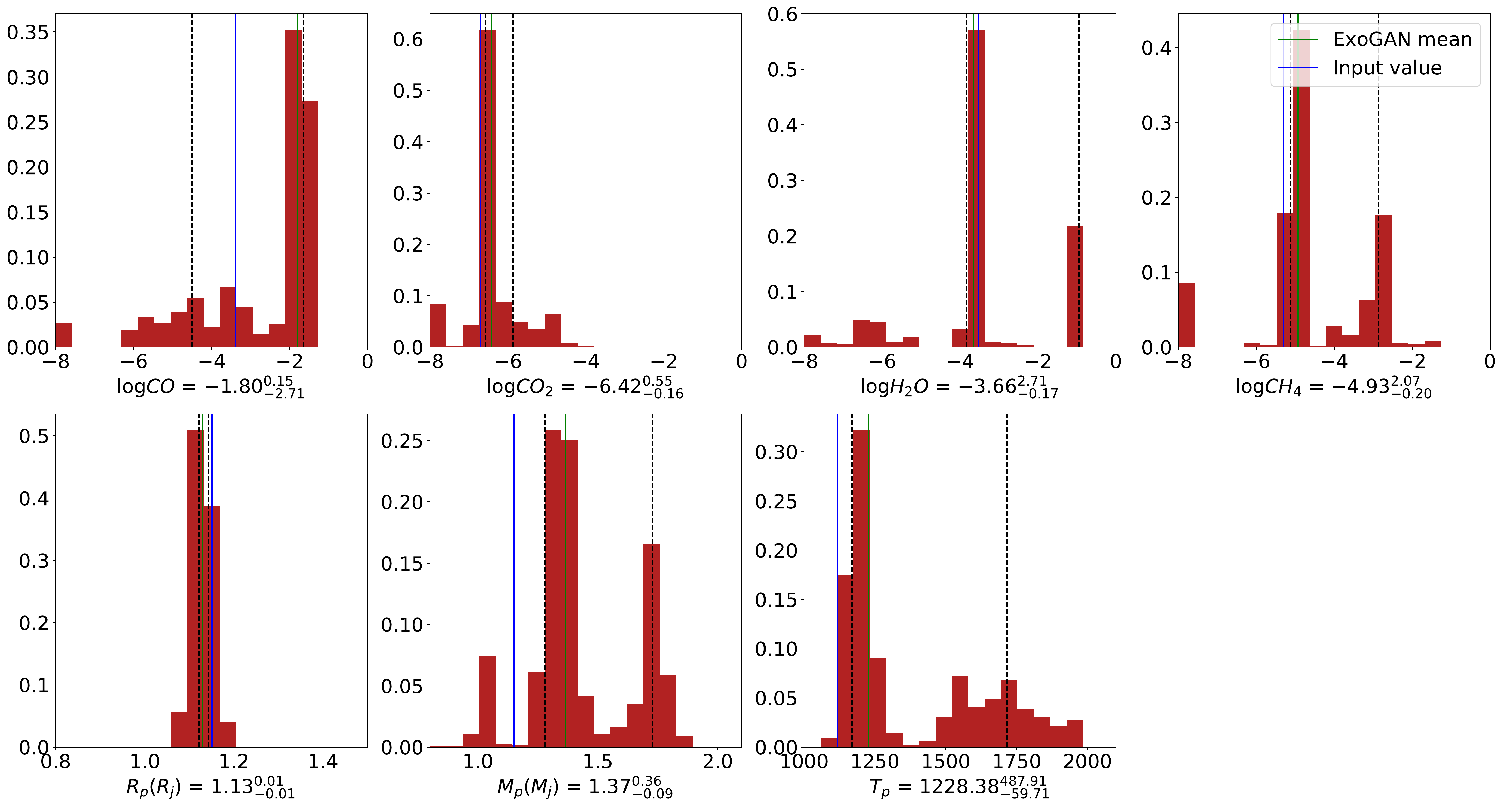}
	\caption{\gan\ parameter distribution of the default test planet. Blue vertical line: Mean predicted value; green vertical line: ground truth value; vertical dotted lines: 1\,$\sigma$ bounds estimated by \gan. \label{fig:iteration_1000}}
\end{figure*}

We compute the reconstruction accuracies for 1000 randomly selected planets for each, the test and training sets. The accuracies are summarised in tables \ref{tab:train_accuracy} \& \ref{tab:accuracy} for 0\,$\sigma$ (an exact match), 1\,$\sigma$ and 2\,$\sigma$ confidence intervals. Figure \ref{fig:iteration_1000} shows an example of the parameter distributions retrieved for a test-case planet.

\begin{table}
	\centering
	\begin{tabular}{ |p{1.5cm}||p{1.5cm}|p{1.5cm}|p{1.5cm}| }
		\hline
		\multicolumn{4}{|c|}{Training set parameters} \\
		\hline
		Variable & $A(0\,\sigma_\phi)$ & $A(1\,\sigma_\phi)$ & $A(2\,\sigma_\phi)$ \\
		\hline
		CO     &  64.4\% & 74.9\%    & 80.8\% \\
		CO$_2$ &  93.7\% & 96.4\%    & 97.3\% \\
		H$_2$O &  86.3\% & 92.9\%    & 94.8\% \\
		CH$_4$ &  80.3\% & 88.4\%    & 91.9\% \\
		R$_p$  &  99.8\% & 99.8\%    & 99.8\% \\
		M$_p$  &  88.8\% & 90.5\%    & 91.6\% \\
		T$_p$  &  89.4\% & 91.9\%    & 93.1\% \\
		\hline
	\end{tabular}
	\caption{\gan\ prediction accuracies associated to each parameters for the training set. The $A(0\,\sigma_\phi)$ column represent the absolute accuracy of the prediction without taking into account the error bar of the retrieval. The 2$^{nd}$ and 3$^{rd}$ columns are taking into account the 1\,$\sigma$ and 2\,$\sigma$ retrieved errors following equation~\ref{eq:accuracy}. \label{tab:train_accuracy}}
\end{table}

\begin{table}
	\centering
	\begin{tabular}{ |p{1.5cm}||p{1.5cm}|p{1.5cm}|p{1.5cm}| }
		\hline
		\multicolumn{4}{|c|}{Test set parameters} \\
		\hline
		Variable & $A(0\,\sigma_\phi)$ & $A(1\,\sigma_\phi)$ & $A(2\,\sigma_\phi)$ \\
		\hline
		CO  &  62.8\%     & 72.6\%    & 78.2\% \\
		CO$_2$ &  94.2\% & 96.6\%    & 97.4\% \\
		H$_2$O &  89.6\% & 92.8\%    & 93.9\% \\
		CH$_4$ &  80.3\% & 88.2\%  & 91.6\% \\
		R$_p$  &  100.0\% & 100.0\%  & 100.0\% \\
		M$_p$  &  88.0\% & 89.7\%    & 90.8\% \\
		T$_p$  &  90.4\% & 92.2\%    & 93.2\% \\
		\hline
	\end{tabular}
	\caption{Same as table~\ref{tab:train_accuracy} but for the test set. \label{tab:accuracy}}
\end{table}

\subsection{Comparison with a classical retrieval model}

\begin{table}
	\centering
	\begin{tabular}{ |c|c| }
		\hline
		\multicolumn{2}{|c|}{Test planet parameters} \\
		\hline
		Parameter & Value \\
		\hline
		$R_{\ast}$    & 0.752\,$R_\odot$ \\
		$R_{p}$    & 1.151\,$R_J$ \\
		$M_{p}$    & 1.150\,$M_J$ \\
		$T_{p}$    & 1117\,K \\
		H$_2$O     & $3\cdot10^{-4}$ \\
		CO     & $4\cdot10^{-4}$ \\
		CO$_2$     & $2\cdot10^{-7}$ \\
		CH$_4$     & $5\cdot10^{-6}$ \\
		\hline
	\end{tabular}
	\caption{Test-case atmospheric and planetary parameters used based on HD\,189733b. The molecular abundances are given in volume mixing ratios. \label{tab:default_parameters}}
\end{table}

In this section, we compare the \gan\ results with a `classical' retrieval result obtained with the TauREx retrieval code. For this comparison and tests in subsequent sections, we used as example the hot-Jupiter HD\,189733b  with planetary/orbital parameters taken from \citet{2008ApJ...677.1324T, 2006ApJ...646..505B} and atmospheric chemistry based on \citet{2012A&A...546A..43V}, see table \ref{tab:default_parameters}.

\begin{figure}
	\centering
	\includegraphics[width=0.5\textwidth]{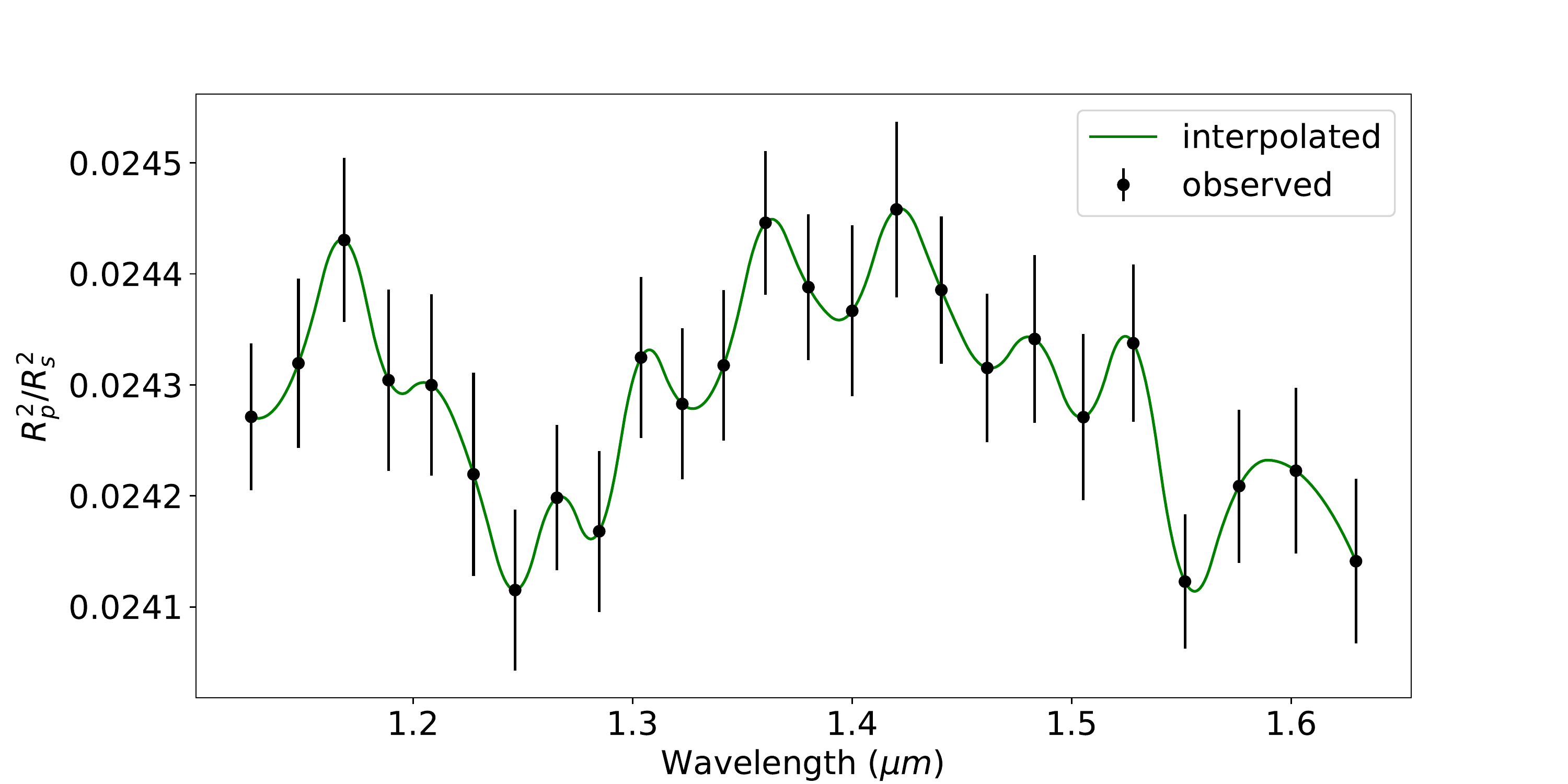}
	\caption{Real HD\,189733b observation with the Hubble WFC3 camera \citep{2018AJ....155..156T}. The black points are the observed data and the green line is the interpolated spectrum to the \gan \ resolution.\label{fig:interp}}
\end{figure}

We now retrieve the forward model parameters for both TauREx and \gan\ for spectra across the Hubble/WFC3 only band and a broad (0.3 - 15\,$\mu$m) wavelength band. 
Here the Hubble/WFC3 spectrum was taken from \citet{2018AJ....155..156T} and interpolated to the \gan\ resolution using a quadratic interpolation (figure \ref{fig:interp}). The large wavelength range spectrum is synthetic, based on table~\ref{tab:default_parameters}. 

\begin{figure}
	\centering
	\includegraphics[width=0.5\textwidth]{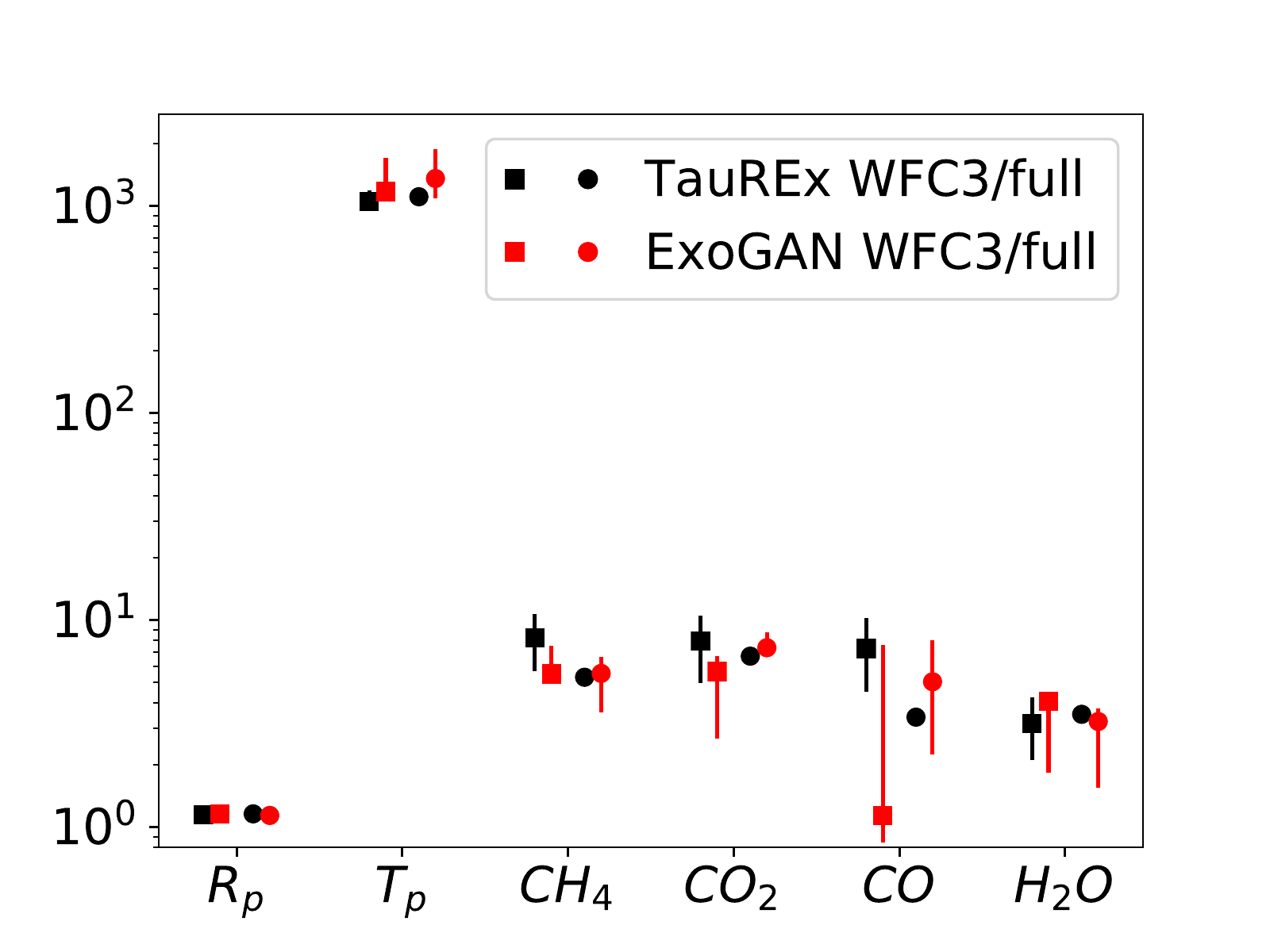}
	\caption{Comparison between the \gan \ predictions (red points) and TauREx (black points). For the molecules we show the value $-log$(mixing-ratio).  The squared points show the results for a real spectrum of HD\,189733b using Hubble/WFC3. The round points are the results for a synthetic model of HD\,189733b between 0.3 - 15\,$\mu$m. The results from the two retrievals are in both cases consistent with each others within the error bars. \label{fig:gan_bayes_comparison}}
\end{figure}

In figure \ref{fig:gan_bayes_comparison} we compare both sets of results. The Hubble/WFC3 and large wavelength retrievals are shown with square and circular markers respectively. In both cases, the \gan\ predictions are consistent with the TauREx retrievals within the error bars.  We note that in the case of CO in the Hubble/WFC3 data, neither TauREx nor \gan\ feature detections as expected.

\begin{figure*}
	\centering
	\includegraphics[width=\textwidth]{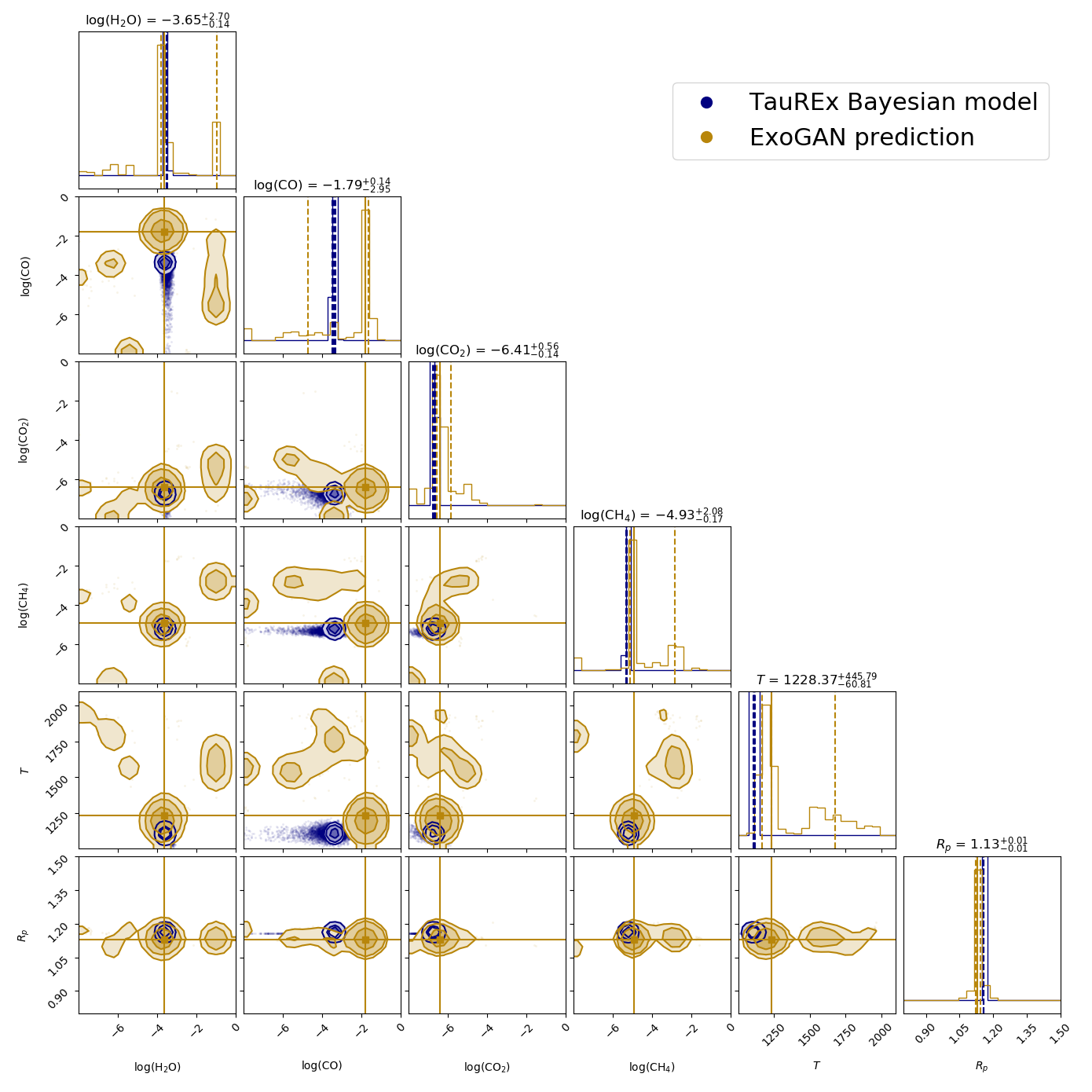}
	\caption{TauREx posterior distributions (in blue) compared to a \gan\ prediction (in golden). As input spectrum, we used a synthetic spectrum of HD\,189733\,b with planetary and atmospheric parameters from \citet{2012A&A...546A..43V} and a wavelength range of 0.3 - 15\,$\mu$m. The two results are in agreement with each other.
\label{fig:full_taurex_exogan}}
\end{figure*}

We then generated a second synthetic spectrum of HD\,189733\,b between $0.3 - 15\,\mu$m, using the parameters of \citet{2012A&A...546A..43V} and overplotted the TauREx retrieved posterior distributions with those derived by \gan, figure~\ref{fig:full_taurex_exogan}. Both algorithms converge to the same solution with the ExoGAN results showing a broader distribution. 
The only significant difference is the CO abundance, where the \gan\ abundances are higher. Note that both, TauREx and \gan\ show tails in their CO abundance posteriors indicating the difficulties of retrieving CO even for classical retrieval algorithms.

Comparisons of run-time are remarkable. Using the TauREx Retrieval code with seven free parameters a standard nested-sampling analysis takes $\sim10$ hours on 24 CPU cores using absorption cross-sections at a resolution of R = 15,000 and spanning a large (0.3 - 15\,$\mu$m) wavelength range. The trained \gan\ requires $\sim 2$ minutes for the same analysis. This result constitutes a speed up of $\sim 300$ times and is independent of the number of free parameters and of the resolution of the input spectrum. Similarly, training \gan\ on higher resolution data, does not significantly impact its runtime after training as both the size and architecture of the underlying network remain unchanged.

\section{Robustness tests}

To test the limits of \gan \ we simulate three conditions previously encountered by the network. We use the same example planet as in the previous section (table \ref{tab:default_parameters}) and simulate the following three scenarios unseen by \gan\ during training phase:

\begin{itemize}
	\item the presence of clouds;
	\item the addition of a trace gas unknown to the network;
	\item atmospheric temperatures outside the training range.
\end{itemize}

\begin{table*}
	\centering
	\begin{tabular}{ |c||c|c|c|c|c|c| }
		\hline
		\multicolumn{7}{|c|}{Robustness results} \\
		\hline
		\multirow{2}*{Variable} & \multicolumn{2}{c|}{clouds} & \multicolumn{2}{c|}{unkwnown gases} & \multicolumn{2}{c|}{$T$ offscale} \\
		\cline{2-7}
		& Input & \gan & Input & \gan & Input & \gan \\
		\hline
		$\log({CO})$               & $-3.4$  & $-4.1_{2.5}^{3.1}$        & $<-8$  & $-5.7_{1.4}^{1.8}$   & $-3.4$ & $-3.1_{3.8}^{0.4}$ \\
		$\log({CO_2})$             & $-6.7$  & $-6.0_{1.7}^{2.3}$        & $<-8$  & $-5.5_{1.8}^{3.9}$   & $-6.7$ & $-5.6_{0.2}^{4.4}$ \\
		$\log({H_2O})$             & $-3.5$  & $-3.6_{3.0}^{1.1}$        & $-3.5$ & $-3.3_{3.5}^{0.7}$   & $-3.5$ & $-2.9_{4.1}^{0.2}$\\
		$\log({CH_4})$             & $-5.3$  & $-6.7_{1.1}^{1.6}$      & $<-8$  & $-5.5_{1.9}^{2.0}$   & $-5.3$ & $-5.1_{1.1}^{2.1}$\\
		R$_p$ (R$_\mathrm{J}$)  & $1.15$  & $1.18_{0.01}^{0.01}$    & $1.15$ & $1.14_{0.01}^{0.01}$ & $1.15$ & $1.16_{0.01}^{0.02}$ \\
		M$_p$ (M$_\mathrm{J}$)  & $1.15$  & $1.23_{0.42}^{0.59}$    & $1.15$ & $1.39_{0.49}^{0.43}$ & $1.15$ & $1.60_{0.7}^{0.2}$\\
		T$_p$ (K)                 & $1117$  & $1681_{208}^{153}$        & $1117$ & $1689_{506}^{179}$   & $2500$ & $1744_{6.4}^{157}$ \\
		\hline
	\end{tabular}
	\caption{Summary of all the robustness test results. For each value we show the input value used for the spectrum and the predicted result from \gan. For the unknown gases test we used ammonia with a volume mixing ratio of $10^{-4}$.\label{tab:robustness_tests}}
\end{table*}

Each test is discussed below, and the \gan\ predicted abundances versus the ground-truth are summarised in table~\ref{tab:robustness_tests}.
Furthermore, we test the \gan's robustness against varying signal-to-noise (S/N) levels of the observed spectrum.

\subsection{Presence of clouds}
\label{sec:clouds}

\begin{figure*}
	\centering
	\includegraphics[width=0.45\textwidth]{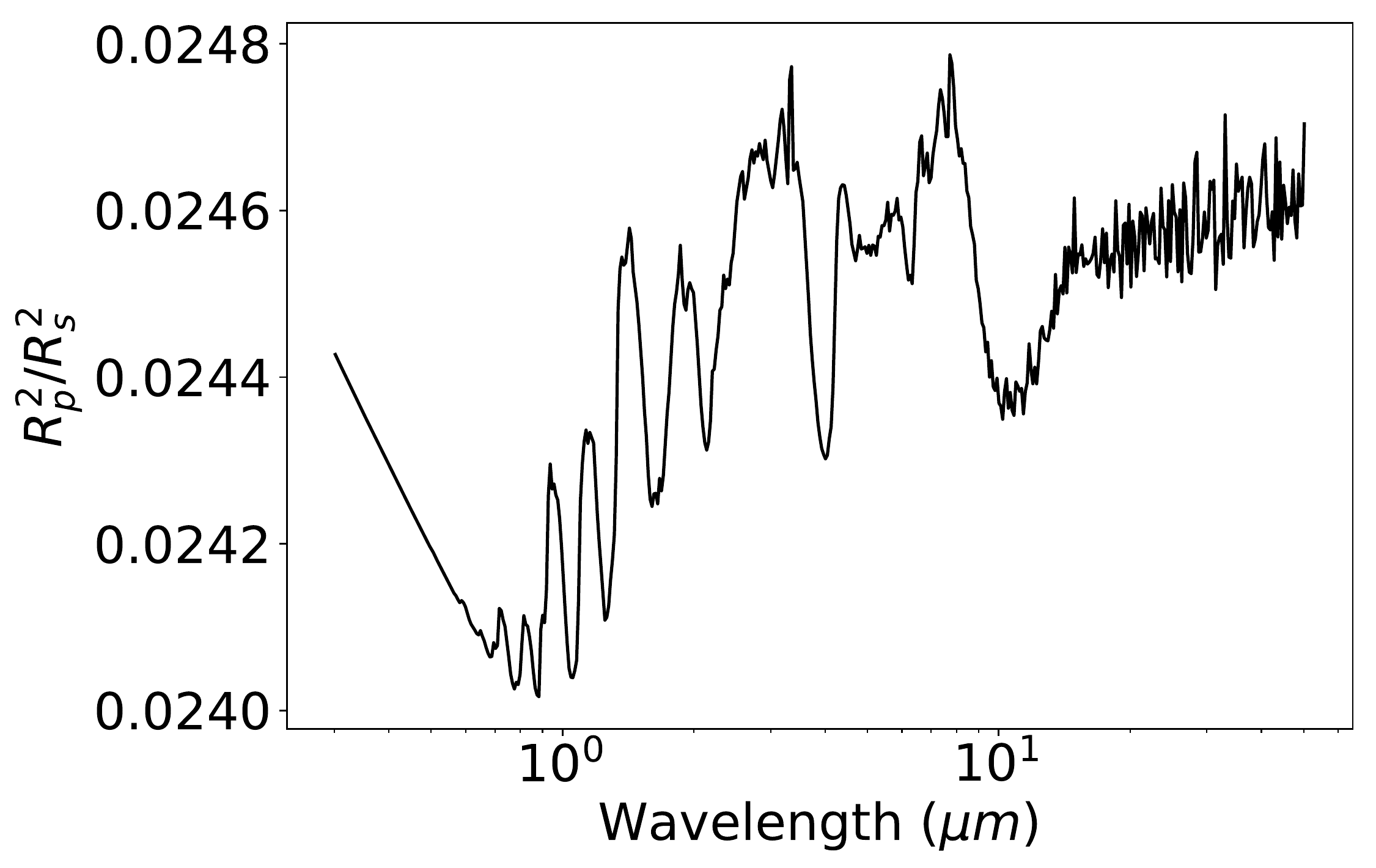}
	\includegraphics[width=0.45\textwidth]{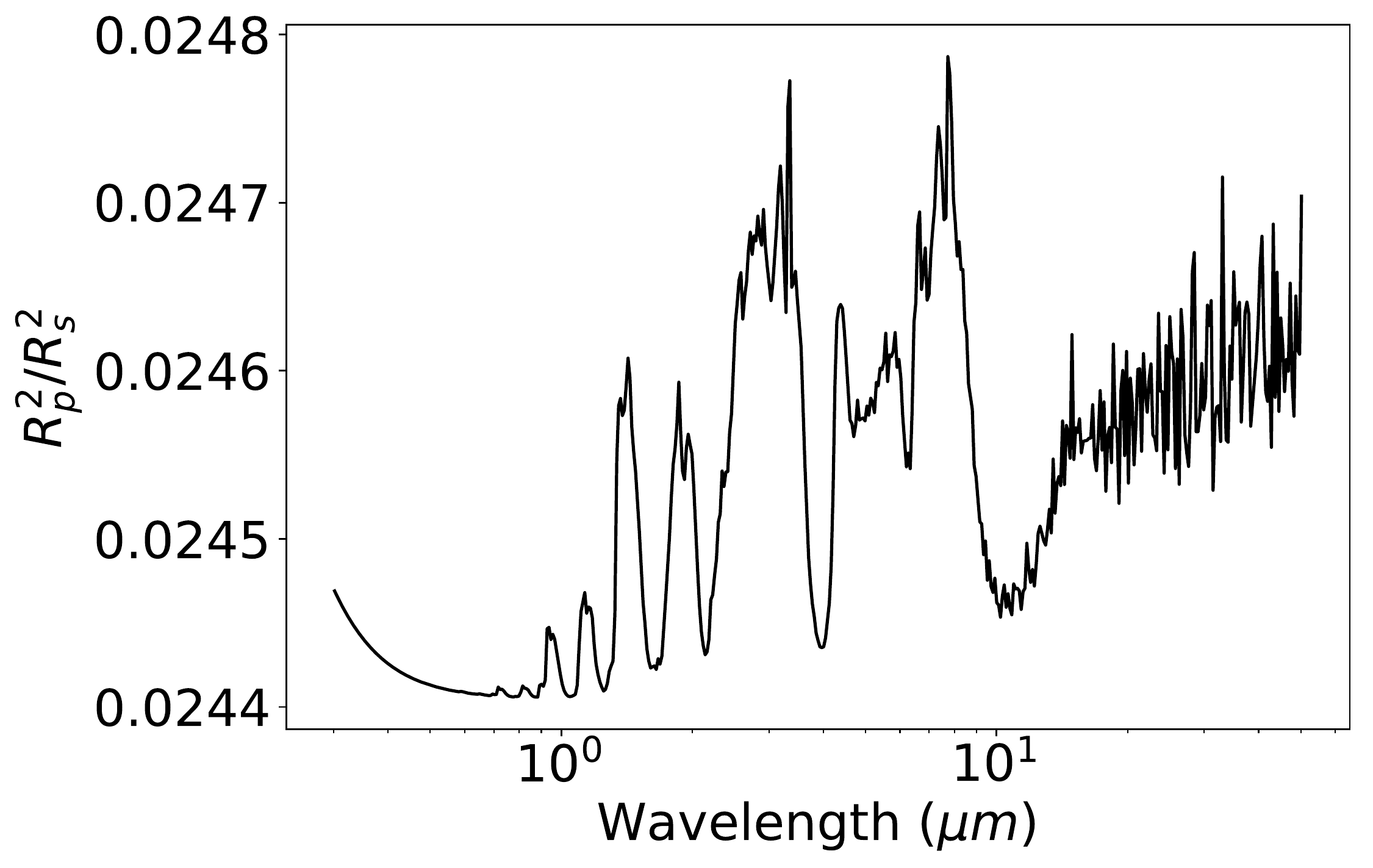}
	\caption{Simulated spectra of the default test planet HD\,189733b without clouds (left) and with grey clouds at 10\,mbar cloud top pressure. (right).\label{fig:clouds}}
\end{figure*}

\begin{figure*}
	\centering
	\includegraphics[width=\textwidth]{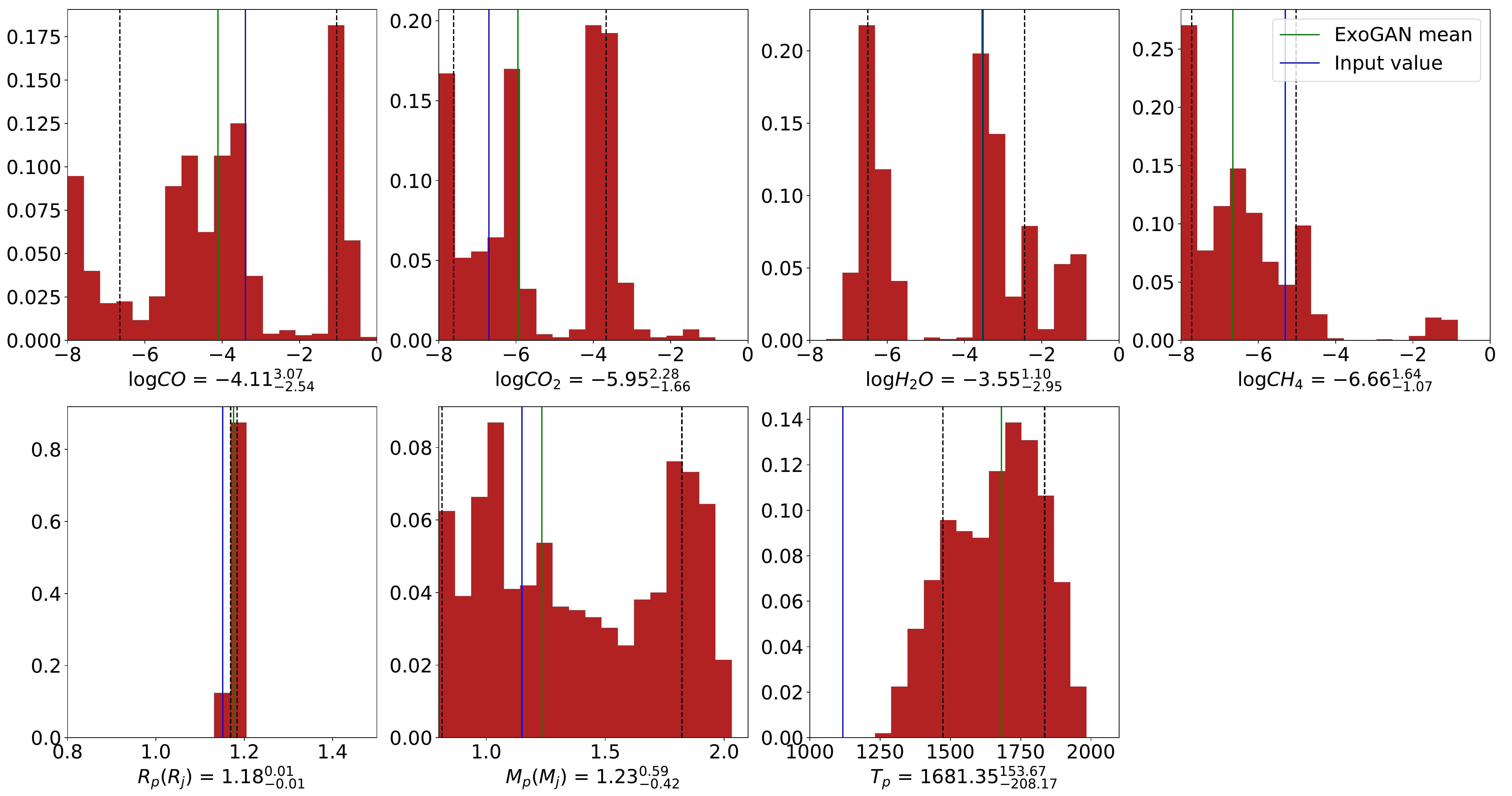}
	\caption{Same as figure \ref{fig:iteration_1000} but for the clouds robustness test for the default test planet, section \ref{sec:clouds}.\label{fig:clouds_1000}}
\end{figure*}

Here we test the response of \gan\ to the presence of clouds in the atmospheric spectrum. 
We simulate a grey cloud deck at 10\,mbar pressure (figure \ref{fig:clouds})
and let \gan \ reconstruct the atmospheric parameters, see figure \ref{fig:clouds_1000}. The lack of information due to the clouds presence results in a wider distribution of parameters. However, \gan \ is still able to retrieve all trace-gas abundances within 1\,$\sigma$ confidence. We find that temperature estimates can be overestimated. This result is likely a consequence of the normalisation procedure used in the presence of clouds. 

\subsection{Presence of molecules outside of the training set}
\label{sec:nh3}
\begin{figure*}
	\centering
	\includegraphics[width=\textwidth]{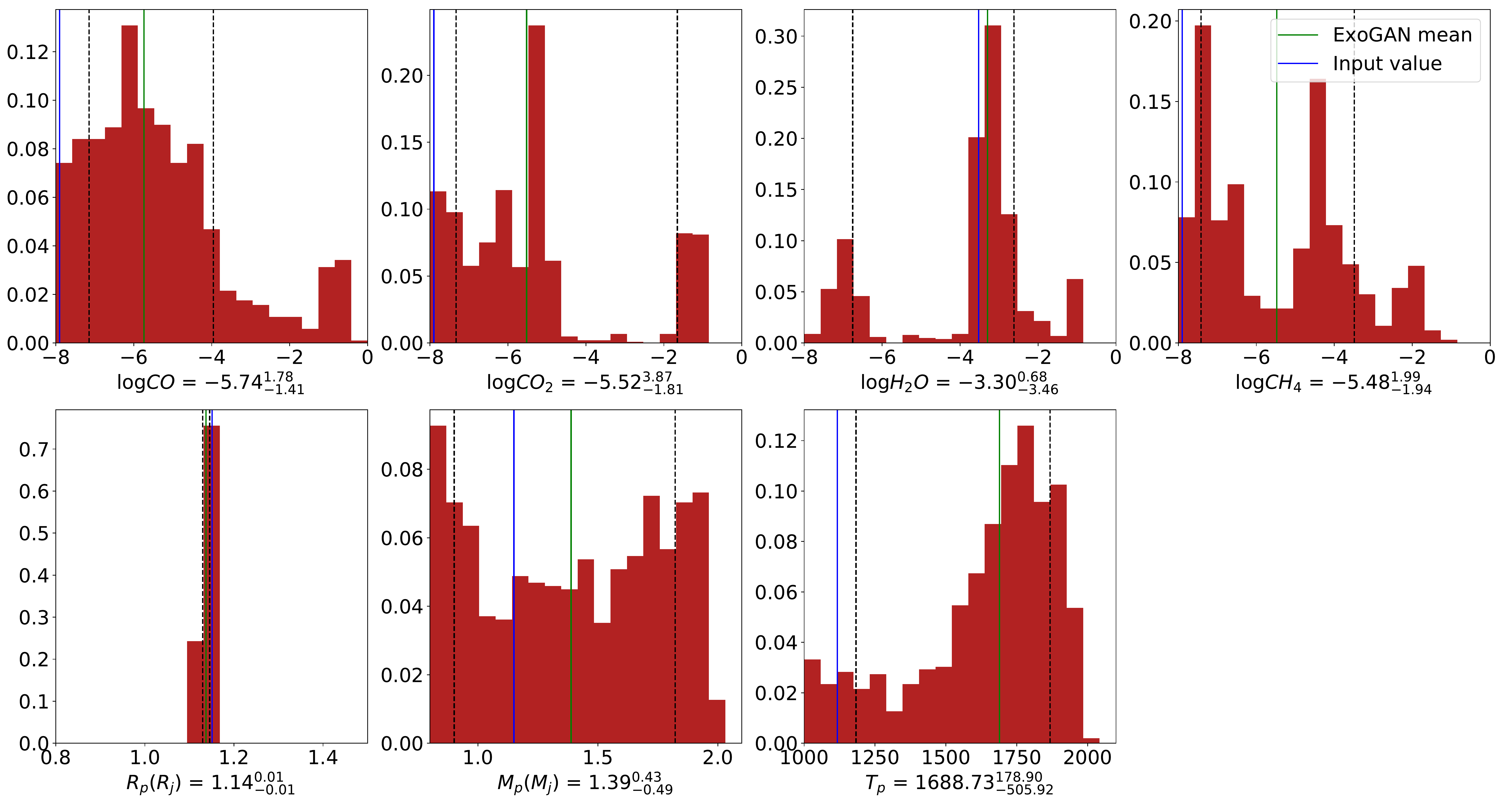}
	\caption{Same as figure \ref{fig:iteration_1000} but for the
		\gan \ analysis for a spectrum with only water and NH$_3$, section \ref{sec:nh3}.\label{fig:nh3_1000}}
\end{figure*}

In this test, we simulate the impact of unknown features on the retrievability of known trace gases. We here consider a spectrum containing water at the default test value and NH$_3$ with a mixing ratio of $10^{-4}$. Though
\citet{2012A&A...546A..43V} estimated an NH$_3$ mixing ratio of $10^{-6}$, we use an unrealistically high value as a worst-case scenario. 
By removing all other trained trace-gases but water, we also test for spurious detections in non-existing trace-gases. Figure \ref{fig:nh3_1000} shows the \gan\ parameter distributions. We find the network to recognise the absence of trace-gases and does not detect `false positives', while still recovering the exact mixing ratio of H$_2$O.

\subsection{Parameters outside the training range}
\label{sec:temperature}

\begin{figure*}
	\centering
	\includegraphics[width=\textwidth]{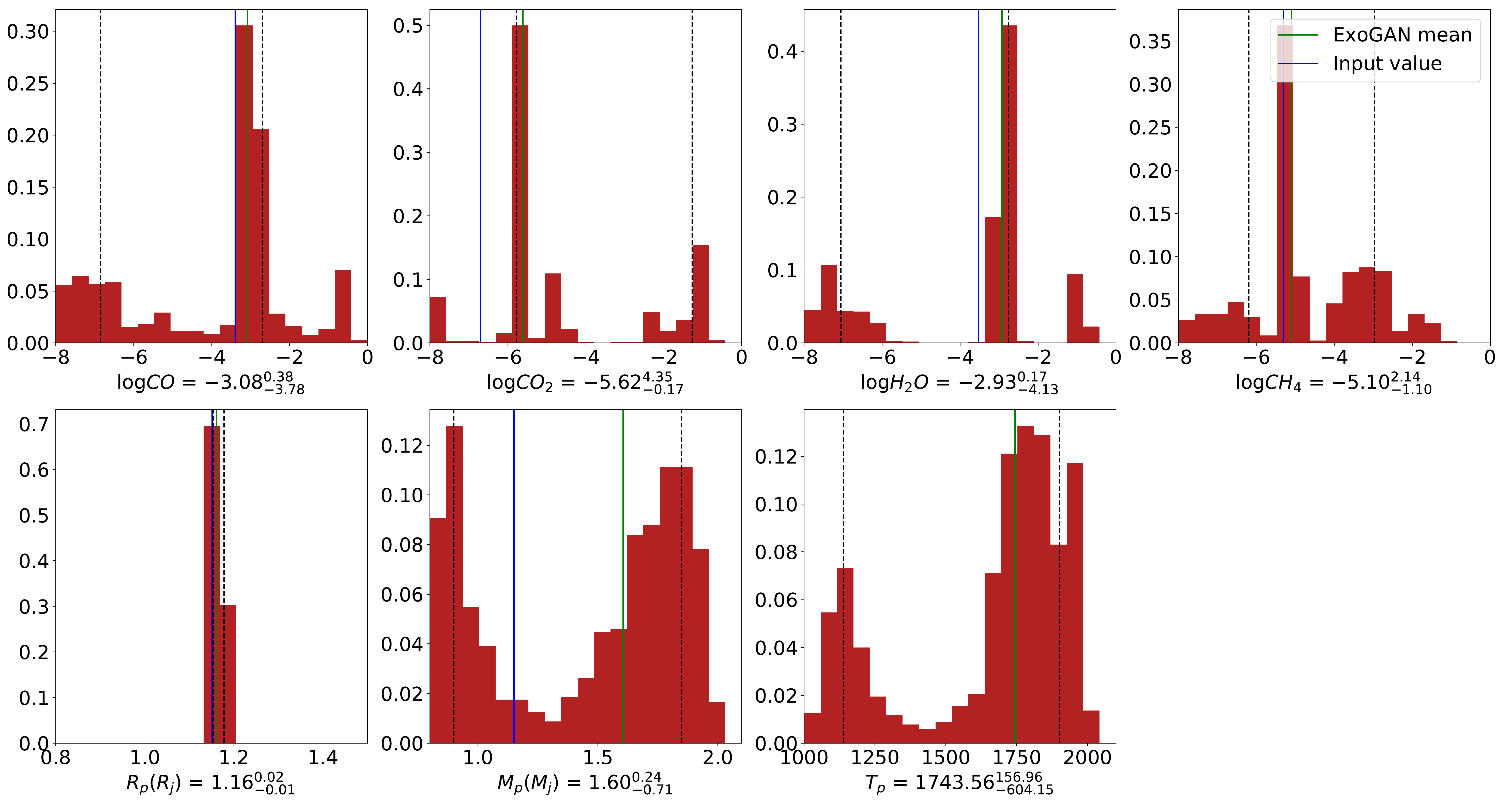}
	\caption{Same as figure \ref{fig:iteration_1000} but for the
		\gan \ analysis for a planetary temperature at 2500\,K, 500\,K outside the training range, see section \ref{sec:temperature}.\label{fig:off_scale_1000}}
\end{figure*}

In the third robustness test we simulated a default planetary atmosphere but an effective temperature of 2500\,K, 500\,K above the temperature training range. In this test, as shown in figure \ref{fig:off_scale_1000}, all parameters converge toward the real solution within 1\,$\sigma$, except for the planetary temperature. Here, the network does not retrieve the correct temperature but assigns a large error bar suggesting that the temperature value is unconstrained if the input value is not contained in the domain range of \gan.

\subsection{Impact of spectral signal-to-noise}

\begin{figure*}
	\centering
	\includegraphics[scale=0.25]{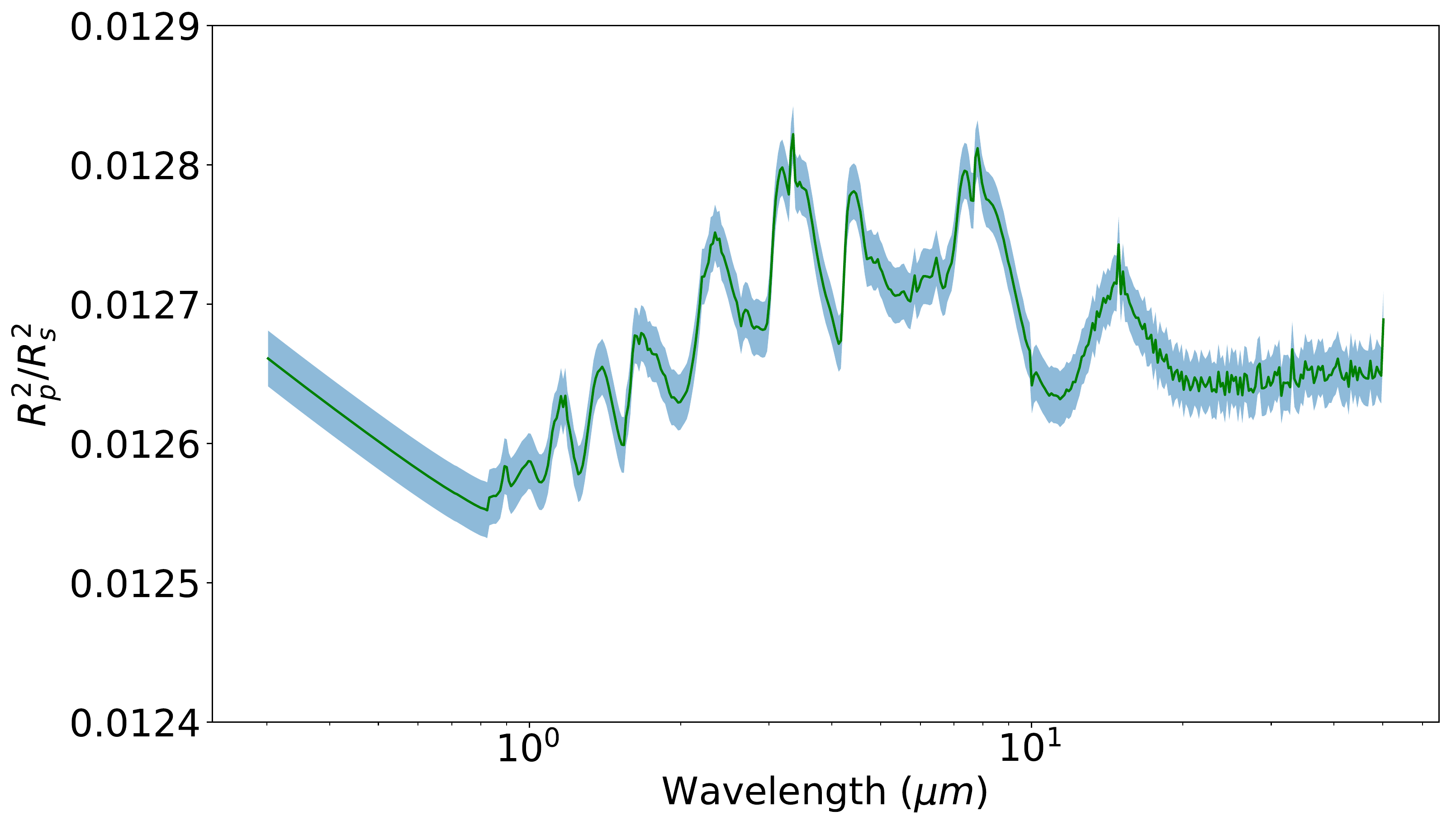}
	\includegraphics[scale=0.25]{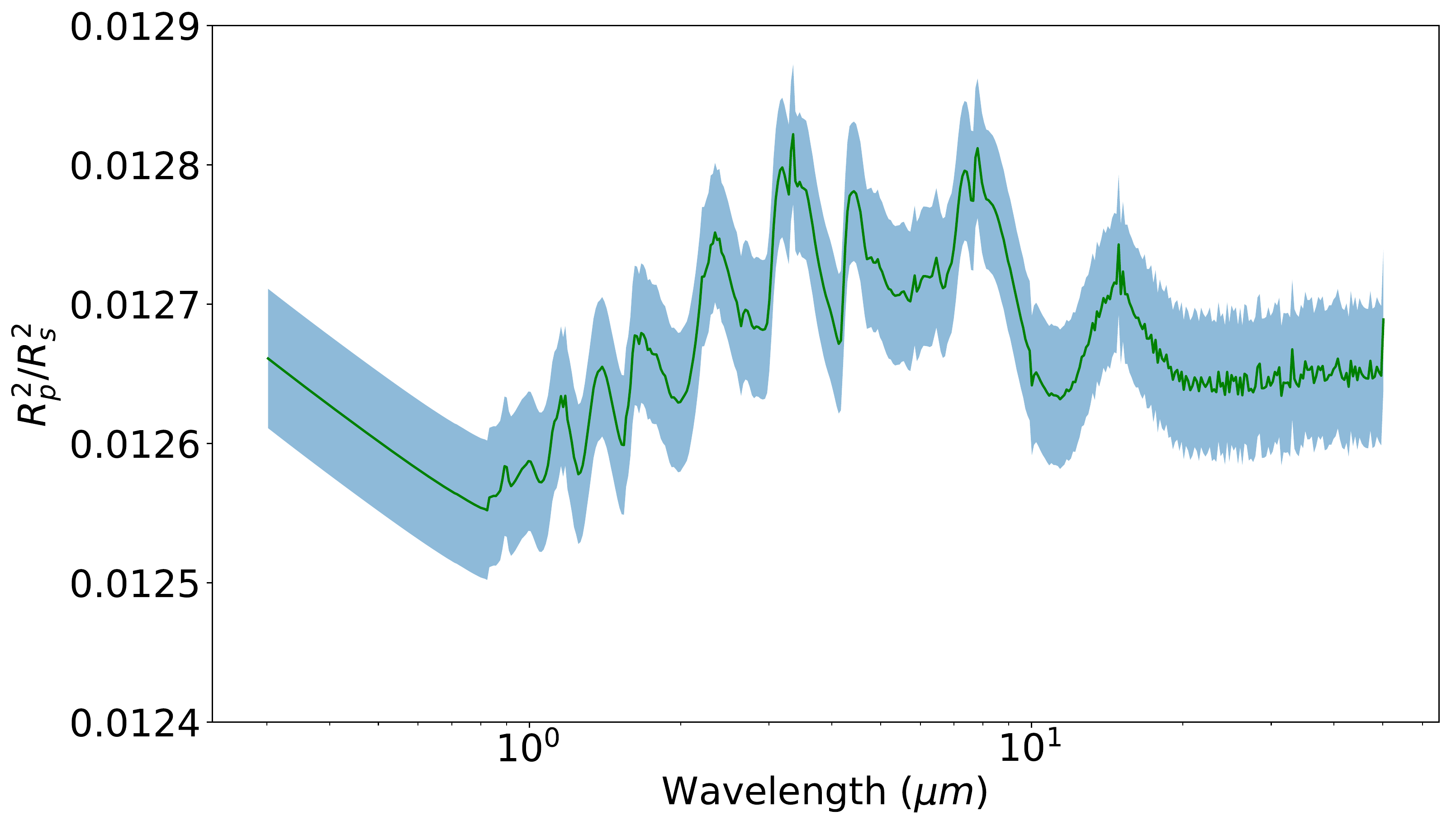}
	\includegraphics[scale=0.25]{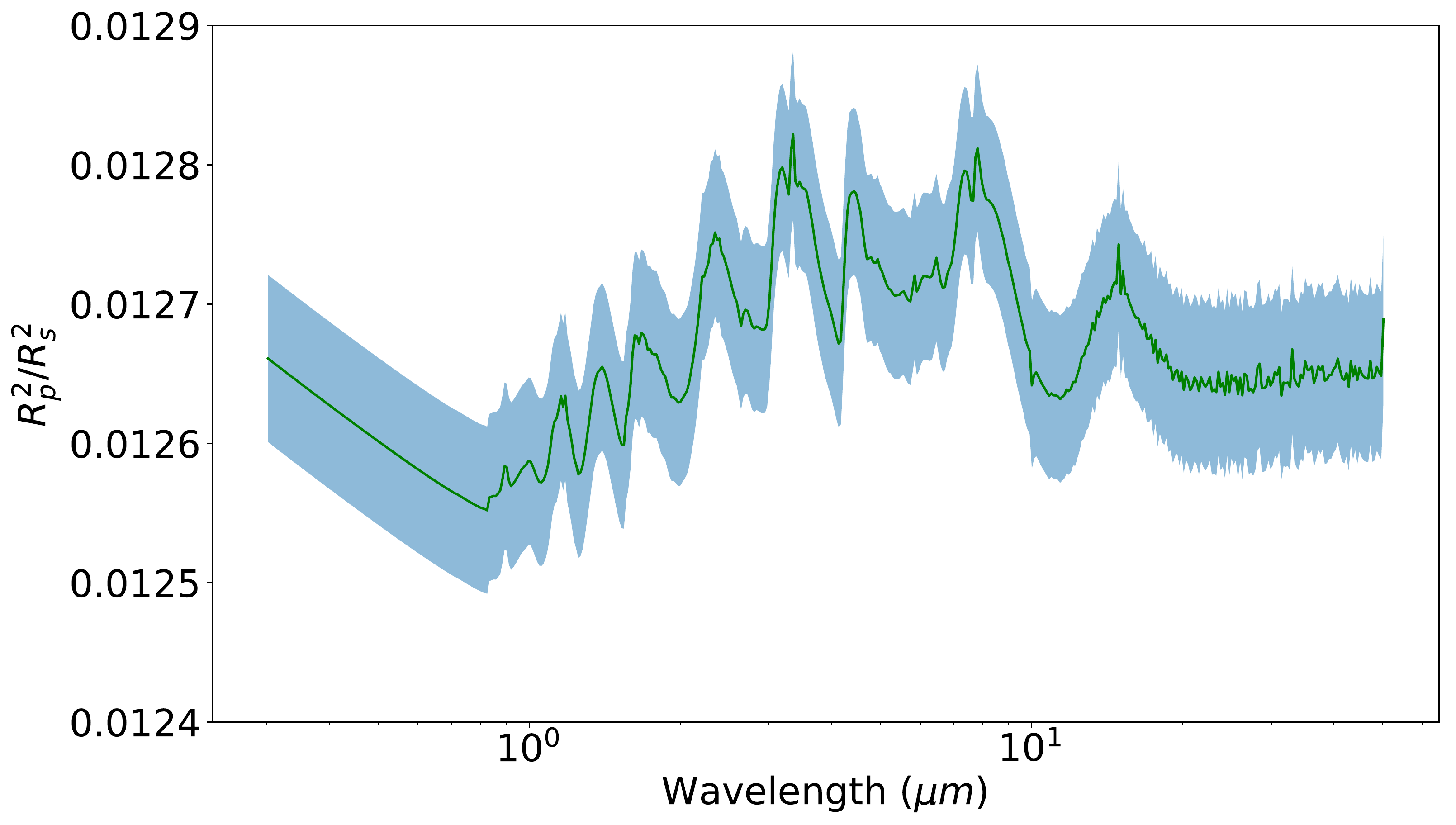}
	\includegraphics[scale=0.25]{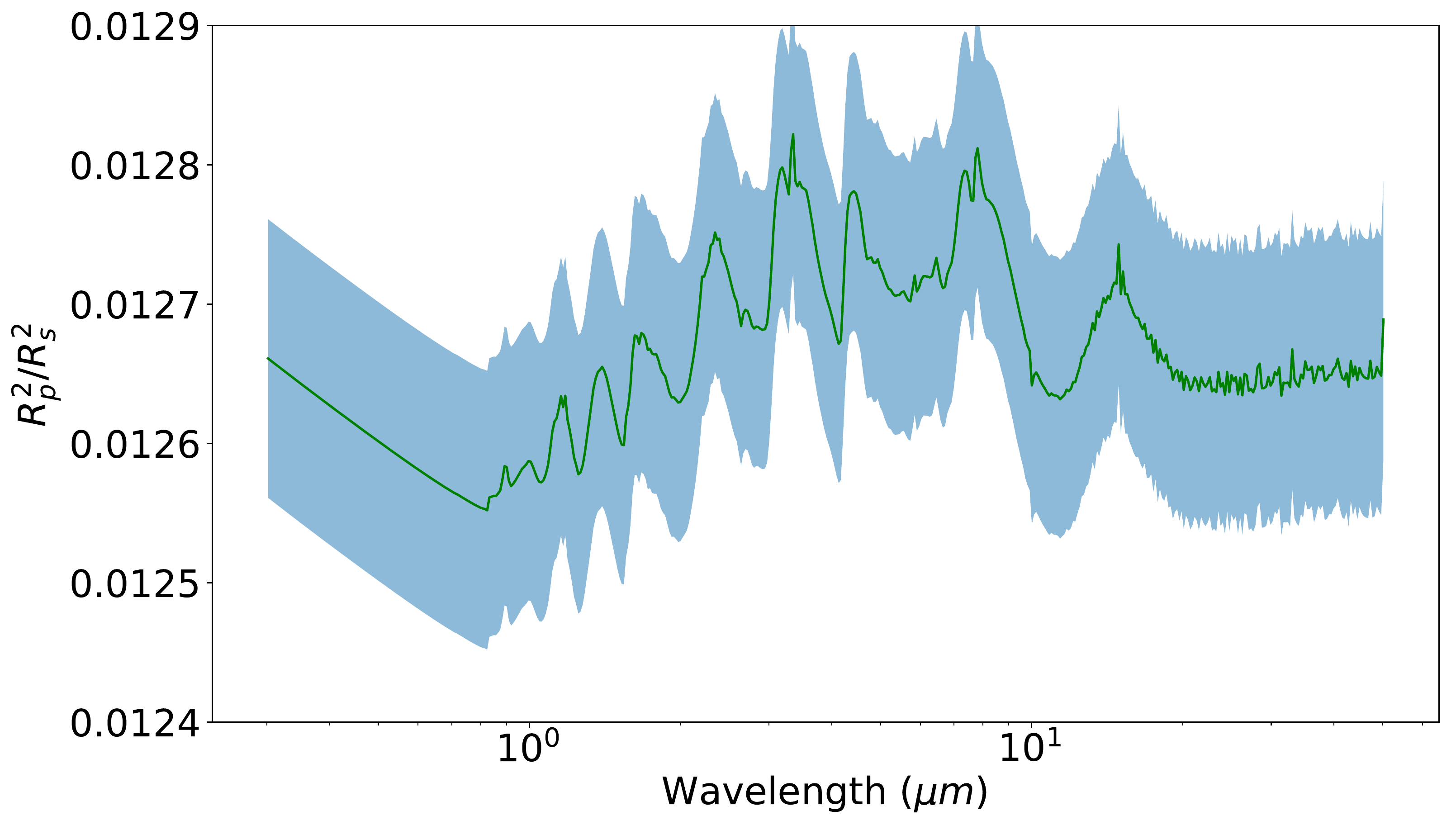}
	\caption{Four examples of spectra used to calculate the accuracy of the \gan. The green line represents the input spectrum and the blue part is the area representing the error bars, $\sigma_\lambda$ in which we varied the input signal to simulate a noisy spectrum. In the \textbf{top left} we seen the 20ppm error bars, in the \textbf{top right} the 50ppm, in the \textbf{bottom left} the 60 ppm and the \textbf{bottom right} the 100ppm one.\label{fig:noisy_spectra}}
\end{figure*}

We test \gan\ for varying levels of observational noise. Here we take the default planet (table~\ref{tab:default_parameters}) and add noise in steps of 10ppm in the range $[0, 100]$ ppm. In figure \ref{fig:noisy_spectra} we show examples of spectra at ${\sigma}_\lambda$: 20, 50, 60 and 100\,ppm noise level. 

\begin{figure}
	\centering

	\includegraphics[width=0.5\textwidth]{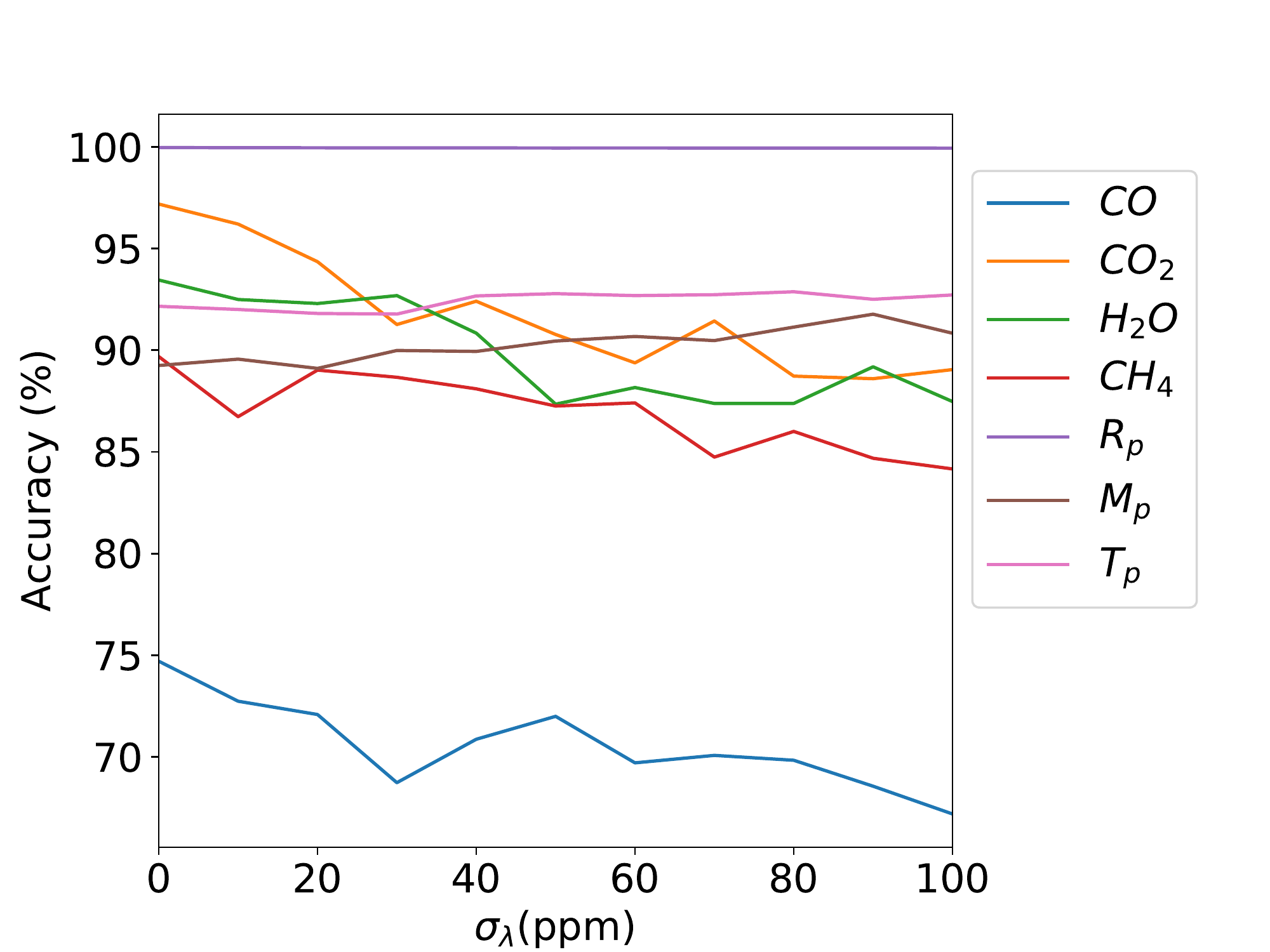}
	\caption{Accuracy as a function of spectral error bars, $\sigma_\lambda$. As discussed in the text, we note that this figure does not take into account the retrieval error bar, i.e. $A(\sigma_\phi=0)$ following equation~\ref{eq:accuracy}. 
		\label{fig:accuracy}}
\end{figure}

For each noise level, we calculated the accuracy of the prediction following equation~\ref{eq:accuracy}, but setting $A(\sigma_\phi = 0)$, figure~\ref{fig:accuracy}. 
We note that figure~\ref{fig:accuracy} only shows the difference between the predicted value and an exact match and prediction accuracies increase when retrieval error bars are taken into account. Here we want to demonstrate the relative degradation of the prediction accuracy as a function of $\sigma_\lambda$.

As intuitively expected, the noisier the spectra, the less accurate the model. The Radius of the planet can be easily recognised by the \gan \ in the entire error range tested. The most difficult parameter to identify is the CO abundance and the mass of the planet.

\section{Discussion}

\subsection{Training and training data}
In this work, we used 10$^7$ forward models over seven atmospheric forward model parameters. We find that this training set is significantly over-complete and the \gan\ training can be completed successfully with $\sim50\,\%$ of the existing training set. Optimising training in future iterations will allow for the inclusion of more complex atmospheric forward models.

One of the main difficulties for training neural networks with transmission spectra is the normalisation of the spectra in $R_p/R_\ast$. A consistent normalisation across a broad range of possible atmospheres is required during the training process, but difficult to achieve in reality given strongly varying atmospheric scale heights and trace-gas abundances. In this work, we adopted a normalisation based on instrument pass-bands as well as water bands. Though in practice this approach works for most scenarios, it can introduce biases when high-altitude clouds are present. In these cases, we find that the normalisation procedure stretches the observed spectrum too much, leading the network to identify higher atmospheric temperatures than it otherwise would. In future work, we plan to mitigate this effect by including grey clouds in the training set as well as further refining the normalisation scheme. We note that for emission spectroscopy a consistent normalisation is more readily achieved if the planetary and stellar equilibrium temperatures are assumed to be known \citep{2016ApJ...820..107W}.  

\gan\ has been trained on a large set of simulated forward models. By including \gan\ as an integral part in the TauREx retrieval framework, we will be able to use forward models created during a standard retrieval run (of the order of 10$^5$ - 10$^6$ models per retrieval) to perform online learning and continuously improve the accuracy of \gan\ over time.

\subsection{Comparison with other machine learning architectures}
In the previous sections, we have explored the use of DCGANs to retrieve atmospheric parameters from observations. GANs belong to the class of semi-supervised and unsupervised generative models and since their inception have been subject to significant research. In this paper, our use of DCGAN is  unsupervised as we provide the parameters together with the data to be modelled. Such an approach allows for a high degree of flexibility in using \gan, as we only need to re-define the ASPA array to train on new problem sets. 

Most other generative models require the likelihood function of the data to be defined, something we do not intrinsically know for many exoplanet observations, whereas GAN based models are likelihood-free methods and  $p_{\bm \theta}({\bf x})$ does not need to be computed during training. This characteristic has obvious advantages over pure variational autoencoders which require a parametrised form of the probability space from which it draws its latent variables. 

Whilst we have explored the use of GANs in the scope of this paper, we note that other neural network architectures, such as simpler deep believe networks or VAEs, may yield comparable results. In fact, recent work by the 2018 NASA Frontier Development Lab\footnote{\href{https://frontierdevelopmentlab.org}{frontierdevelopmentlab.org}}
 has explored various deep learning architecture in the context of atmospheric retrievals with promising results. 
Similarly, other machine learning frameworks may also be successfully used to model exoplanetary spectra. For example, \citet{2018NatAs.tmp...81M} recently  presented an atmospheric retrieval algorithm based on random forests regression \citep{Breiman2001} and demonstrated the algorithm on Hubble/WFC3 observations.

\subsection{Future work}

In this work, we have used the `vanilla' DCGAN as underlying algorithm. Since its inception, various interesting additions to the classical GAN have been proposed which we intend to explore in future work. Notable amongst them are the VAE-GAN hybrids, random forest and GAN hybrids and Bayesian-GAN models. 
The VAE-GAN models \citep[e.g.][]{2017arXiv170604987R,NIPS2016_6158,2017arXiv170402304U,2015arXiv151105644M}, allow direct inference using GANs. Something that is not possible using purely generative models. 
To further guard against model collapse, \citet{2018arXiv180505185Z} have recently proposed a random forest and GAN hybrid algorithm, GAF, where the fully connected layer of the GAN's discriminator is replace by a random forest classifier. 
\citet{2017arXiv170509558S} proposed a Bayesian-GAN by drawing probability distributions over ${\bm \theta}^{(D)}$ and  ${\bm \theta}^{(G)}$, allowing for fully bayesian predictive models and further guarding against model collapse.

\section{Conclusion}

In the era of JWST and ARIEL observations, next-generation atmospheric retrieval algorithms must reflect the higher information content of the observation with an increase in atmospheric model complexity. Complex models are computationally heavy, creating potential bottlenecks given current state-of-the-art sampling schemes. 
Artificial intelligence approaches will provide essential tools to mitigate the increase in computational burden while maintaining retrieval accuracies. 

In this work, we introduced the first deep learning approach to solving the inverse retrieval of exoplanetary atmospheres. We trained deep convolutional generative adversarial network on an extensive library of atmospheric forward models and their associated model parameters. The training set spans a broad range of atmospheric chemistries and planet types. Once trained, the \gan\ algorithm achieves comparable performances to more traditional statistical sampling based retrievals, and the \gan\ results can be used to constrain the prior ranges of subsequent retrievals (to significantly cut computation times) or be used as stand-alone results. We found \gan\ to be up to 300 times faster than a standard retrieval for large spectral ranges. \gan\ is designed to be universally applicable to a wide range of instruments and wavelength ranges without additional training.

\begin{acknowledgements}
	All codes used in this publication are open-access and their latest versions are hosted at \url{https://github.com/orgs/ucl-exoplanets}. Manuals and links to the training sets can be found at \url{http://exoai.eu}.\\
Furthermore, the training data and the corresponding ExoGAN software (at the time of paper acceptance) have been assigned the DOI:10.17605/OSF.IO/6DXPS and are permanently archived at \url{https://osf.io/6dxps/}. All data/software pertaining to the ExoAI project (inc. TauREx) is archived here: \url{https://osf.io/tfyn6/}
	
	The authors thank the anonymous referee for improving the clarity of the paper. The authors also thank the UCL Exoplanet team for insightful discussions. 
	This project has received funding from the European Research Council (ERC) under the European Union's Horizon 2020 research and innovation programmes (grant agreement No 758892/ExoAI and and No 776403/ExoplANETS A) and the European Union's Seventh Framework Programme (FP7/2007-2013) ERC grant agreement No 617119/ExoLights. The authors further acknowledge funding from Microsoft Azure for Research and the STFC grants ST/K502406/1 and ST/P000282/1. TZ also acknowledges the contribution from INAF trough the ``Progetti Premiali'' funding scheme of the Italian Ministry of Education, University, and Research
\end{acknowledgements}

\bibliographystyle{apj} 
\bibliography{bibliography}

\appendix

\section{The ADAM optimiser} \label{app:adam}

The Adaptive Moment Estimation (ADAM) is a very popular algorithm in deep learning and it computes adaptive learning rate for the parameters of a neural network. It stores the exponentially decaying average of past squared gradients $v_t$ together with the exponentially decaying average of the past gradients $m_t$. Keeping the notation of \citet{2016arXiv160904747R}, the past and the squared past gradients, $m_t$ and $v_t$ are defined as:

\begin{equation}
m_t = \beta_1 m_{t-1} + (1-\beta_1)g_t
\label{eq:gradient}
\end{equation}

and,

\begin{equation}
v_t = \beta_2 v_{t-1} + (1-\beta_2)g_t^2,
\label{eq:squared_gradient}
\end{equation}

with $\beta_1$ and $\beta_2$ being the decay rates, and $g_t = \nabla_{\bf{z}} \mathcal{L}(\bf{z}_t)$ the gradient of the $\mathcal{L}$ function defined in Equation \ref{eq:complete_loss}

Equations \ref{eq:gradient} and \ref{eq:squared_gradient} estimates, respectively, the mean (or first moment) and the variance (or second moment) of the gradients. Since the two moments are initialised as vectors of 0's, they are biased towards zero, particularly during the first time steps or when the decay rates $\beta_1$ and $\beta_2$ are small. To correct the biases \citet{2014arXiv1412.6980K} defined the bias-corrected moments as:

\begin{equation}
\hat{m}_t = \frac{m_t}{1-\beta_1^t}
\label{eq:first_bias_corrected}
\end{equation}

and,

\begin{equation}
\hat{v}_t = \frac{v_t}{1 - \beta_2^t}.
\label{eq:second_bias_corrected}
\end{equation}

At this point it is possible to update the $\bf{z}$ variable using the Adam update rule:

\begin{equation}
\textbf{z}_{t+1} = \textbf{z}_t - \frac{\eta}{\sqrt{\hat{v}_t} + \varepsilon} \hat{m}_t
\label{eq:adam_rule}
\end{equation}

We used the values suggested by \citet{2014arXiv1412.6980K} for the hyperparameters, shown in Table \ref{tab:gan_params}.


\section{Batch Normalisation} \label{app:bn}

A characteristics of DCGANs is the use of batch normalisation (BN) \citep{2015arXiv150203167I, 2017arXiv170403971X}. BN is now a common technique in deep learning applications to accelerate the training of neural networks. DCGAN networks \citep{2015arXiv151106434R} use BN for both the Discriminator and the Generator nets. Nevertheless, GAN architectures started using BN just for the generator net with the LAPGAN networks \citep{2015arXiv150605751D}. Nowadays, many GAN architectures use BN. The idea behind BN is using a batch of samples $\left\lbrace x_1, x_2, ..., x_m \right\rbrace$ and computing, keeping the notation of \citet{2017arXiv170403971X}, the following:

\begin{equation}
y_i = \frac{x_i - \mu_B}{\sigma_B}\cdot\gamma + \beta,
\label{eq:BN}
\end{equation}

with $\mu_B$, $\sigma_B$, respectively, the mean and the standard deviation of the batch and $\gamma$ and $\beta$ the learned parameters. BN allows to have an output with a mean $\mu$ and a standard deviation $\sigma$ independently on the input distribution.

\section{\gan \ architecture and parameters}

\gan \ is made up of two neural networks, the generator and the discriminator, whose parameters are shown in Tab \ref{tab:gan_architecture}.

\begin{table}[!htbp]
	\centering
	\begin{tabular}{ cccc }
		\hline
		Layer & Operation & Output & Dimension\\
		\hline
		\multicolumn{2}{c}{\it{Discriminator (${\bm \theta}^{(D)}$)}} & &  \\
		\hline
		\textbf{X} & & & $m \cdot 33 \cdot 33 \cdot 1$\\
		$h_0$ & conv & leaky relu - batch norm & $m \cdot 17 \cdot 17 \cdot 64$ \\
		$h_1$ & conv & leaky relu - batch norm & $m \cdot 9 \cdot 9 \cdot 128$\\
		$h_2$ & conv & leaky relu - batch norm & $m \cdot 5 \cdot 5 \cdot 256$\\
		$h_3$ & conv & leaky relu - batch norm & $m \cdot 3 \cdot 3 \cdot 512$\\
		$h_4$ & linear& sigmoid & $m \cdot 1$\\
		\hline
		\multicolumn{2}{c}{\it{Generator (${\bm \theta}^{(G)}$)}} & &  \\
		\hline
		\textbf{z} & & & $m \cdot 100$\\
		$h_0$ & linear & relu - batch norm & $m \cdot 3 \cdot 3 \cdot 512$ \\
		$h_1$ & deconv & relu - batch norm & $m \cdot 5 \cdot 5 \cdot 256$\\
		$h_2$ & deconv & relu - batch norm & $m \cdot 9 \cdot 9 \cdot 128$\\
		$h_3$ & deconv & relu - batch norm & $m \cdot 17 \cdot 17 \cdot 64$\\
		$h_4$ & deconv & sigmoid & $m \cdot 33 \cdot 33 \cdot 1$\\
		\hline
	\end{tabular}
	\caption{Architecture of \gan\ listing the hyperparameters ${\bm \theta}^{(D)}$ and  ${\bm \theta}^{(G)}$. We used 5 layer deep networks for both Generator and Discriminators. $m$ is the batch size fixed to 64 during training.\label{tab:gan_architecture}}
\end{table}

\begin{table}[!htbp]
	\centering
	\begin{tabular}{ cccp{7cm} }
		\hline
		\multirow{2}*{Hyper-parameter} & \multicolumn{2}{c}{\it{Stage}} & \multirow{2}*{Description}\\
		& Training & Prediction & \\
		\hline
		\multirow{2}*{batch size} & \multirow{2}*{64} & \multirow{2}*{1024}& Number of spectral samples used at each training/prediction iteration for both networks\\
		& & & \\
		
		\textbf{z} & 100 & 100& Generator gaussian prior distribution\\
		$\eta$ & $2\cdot 10^{-4}$& $1\cdot 10^{-1}$ & Learning rate for the Adam optimizer \\
		\multirow{2}*{$\beta_1$} & \multirow{2}*{$0.5$} & \multirow{2}*{$0.9$} & Exponential decay rate for the first moment estimates in the Adam optimizer. \\
		& & & \\
		\multirow{2}*{$\beta_2$} & \multirow{2}*{-} & \multirow{2}*{$0.999$} & Exponential decay rate for the second moment estimates in the Adam optimizer. \\
		& & & \\
		\multirow{3}*{$\lambda$} & \multirow{3}*{-} & \multirow{3}*{$0.1$} & Hyper-parameter that controls the importance of the contextual loss compared to the perceptual loss \\
		
        \multirow{3}*{$\varepsilon$} & \multirow{3}*{-} & \multirow{3}*{$10^{-8}$} & Constant which prevents the denominator in Equation \ref{eq:adam_rule} to be zero \\
        
		\hline
		
	\end{tabular}
	\caption{Hyperparameters used in \gan.\label{tab:gan_params}}
\end{table}

\end{document}